\documentclass[11pt]{article}

\usepackage[dvips]{color}
\usepackage{epsfig}
\usepackage{amsmath}
\usepackage{amsfonts}
\usepackage{smartref} 

\usepackage{fullpage}

\newcommand{\later}[1]{}

\newcommand{\capspace}{\vspace*{-5mm}}

\newcounter{linenum}
\addtoreflist{linenum}
\def\codeTabSpace{\hspace*{6mm}}
\newenvironment{code}%
{\begin{tabbing}%
\codeTabSpace \= \hspace*{35mm} \= \hspace*{83mm} \= \kill%
}%
{\end{tabbing}%
}
\newcounter{ind}
\newcommand{\n}{\addtocounter{ind}{7}\hspace*{7mm}}
\newcommand{\p}{\addtocounter{ind}{-7}\hspace*{-7mm}}
\newcommand{\nl}{\\[-.2mm]\stepcounter{linenum}{\scriptsize \arabic{linenum}}\>\hspace*{\value{ind}mm}}
\newcommand{\ul}{\\[-.2mm]\>\hspace*{\value{ind}mm}}
\newcommand{\bl}{\\[-1.5mm]\>\hspace*{\value{ind}mm}}
\newcommand{\firstline}{\stepcounter{linenum}{\scriptsize \arabic{linenum}}\>}
\newcommand{\lref}[1]{\linenumref{#1}} 

\newcommand{\record}{record}

\newcommand{\flag}{\mbox{\sc Flag}}
\newcommand{\commit}{\mbox{\sc Commit}}
\newcommand{\abort}{\mbox{\sc Abort}}
\newcommand{\try}{\mbox{\sc Try}}
\newcommand{\Scan}{\mbox{\sc RangeScan}}
\newcommand{\ScanHelper}{\mbox{\sc ScanHelper}}
\newcommand{\Insert}{\mbox{\sc Insert}}
\newcommand{\Delete}{\mbox{\sc Delete}}
\newcommand{\Find}{\mbox{\sc Find}}
\newcommand{\Search}{\mbox{\sc Search}}
\newcommand{\Help}{\mbox{\sc Help}}
\newcommand{\Execute}{\mbox{\sc Execute}}
\newcommand{\ReadChild}{\mbox{\sc ReadChild}}
\newcommand{\ValidateLink}{\mbox{\sc ValidateLink}}
\newcommand{\ValidateLeaf}{\mbox{\sc ValidateLeaf}}

\newcommand{\Frozen}{\mbox{\sc Frozen}}
\newcommand{\FScan}{\mbox{\sc Scan}}

\newcommand{\NBBST}{\mbox{\sf NB-BST}}
\newcommand{\NBBSTRQ}{\mbox{\sf PNB-BST}}

\newcommand{\Key}{\mbox{Key}}

\newcommand{\NULL}{\mbox{$\bot$}}
\newcommand{\clean}{\mbox{\sc Clean}}
\newcommand{\mk}{\mbox{\sc Mark}}

\newcommand{\Info}{Info}
\newcommand{\Flag}{Flag}
\newcommand{\DFlag}{D\Flag}
\newcommand{\IFlag}{I\Flag}

\newcommand{\ie}{{\it i.e.}}

\newcommand{\y}[1]{{\color{red} #1}\normalcolor}

\newcommand{\func}[1]{\mbox{\sc #1}}

\newcommand{\CAS}{\func{CAS}}
\newcommand{\LLX}{\func{LLX}}
\newcommand{\SCX}{\func{SCX}}
\newcommand{\VLX}{\func{VLX}}

\newcommand{\comnospace}{\mbox{$\triangleright$}}
\newcommand{\com}{\mbox{\comnospace\ }}

\newtheorem{theorem}{Theorem}

\newtheorem{corollary}[theorem]{Corollary}
\newtheorem{lemma}[theorem]{Lemma}

\newtheorem{definition}[theorem]{Definition}
\newtheorem{observation}[theorem]{Observation}

\newtheorem{invariant}[theorem]{Invariant}
\newcommand{\qedsymb}{\hfill{\rule{2mm}{2mm}}}
\newenvironment{proof}{\begin{trivlist}
\item[\hspace{\labelsep}{\bf\noindent Proof: }]
}{\qedsymb\end{trivlist}}

\newcommand{\remove}[1]{}
\newcommand{\ignore}[1]{}

\newcommand{\lft}{\mbox{\it left}}
\newcommand{\LEFT}{\mbox{\sc left}}
\newcommand{\RIGHT}{\mbox{\sc right}}
\newcommand{\TRUE}{\mbox{\sc True}}
\newcommand{\FALSE}{\mbox{\sc False}}

\newcommand{\tabtabcom}{\>\>\com}
\newcommand{\tabcom}{\>\com}

\date{FORTH ICS TR 470, May 2018}

\begin{document}

\title{Persistent Non-Blocking Binary Search Trees\\Supporting Wait-Free Range Queries}

\author{Panagiota Fatourou\\FORTH ICS \& University of Crete\\Greece 
\and Eric Ruppert\\York University\\Canada}

\maketitle
\thispagestyle{empty}

\begin{abstract}
This paper presents the first implementation of a search tree data structure 
in an asynchronous shared-memory system that provides a {\em wait-free}
algorithm for executing range queries on the tree, 
in addition to non-blocking algorithms for \Insert, \Delete\ and \Find,
\y{using single-word Compare-and-Swap (\CAS). 
The implementation is linearizable
and tolerates any number of crash failures. }
\Insert\ and \Delete\
operations that operate on different parts of the tree
run fully in parallel (without any interference with one another).
We employ a lightweight helping mechanism, where each \Insert, \Delete\ and \Find\ operation
helps only update operations that affect the local neighbourhood of the
leaf it arrives at. 
Similarly, a \Scan\ helps only those updates 
taking place on nodes of the part of the tree it traverses,
and therefore \Scan s operating on different parts 
of the tree do not interfere with one another.
Our implementation works in a dynamic system 
where the number of processes may change over time.  

The implementation builds upon the non-blocking binary search tree
implementation presented by Ellen {\em et al.} \cite{EFRB10} 
by applying a simple mechanism to make the tree persistent. 
\end{abstract}









\section{Introduction}

There has been much recent work
on designing efficient concurrent implementations of {\em set} data structures~\cite{BP12,BCCO10,BER14,CNT14,EFHR14,EFRB10,HJ12,NM14,PBBO12,Sha13},
which provide algorithms
for \Insert, \Delete, and \Find. 
There is increasing interest in providing additional operations for modern applications,
including iterators~\cite{ALRS17,NGPT15b,NGPT15a,PT13,PBBO12,Ros} 
or general range queries~\cite{BA12,Cha17}. These are required 
in many  big-data applications~\cite{CGZ+17,KNT00,P15},
where shared in-memory tree-based data indices must be created 
for fast data retrieval and useful data analytics.
Prevalent programming frameworks (e.g., Java~\cite{javase7}, .NET~\cite{dotNET45}, TBB~\cite{intelTBB})
that provide 
concurrent data structures have 
added operations to support
(non-linearizable)~iterators.

The Binary Search Tree (BST) is one of the most fundamental data structures.
Ellen {\em et al.}~\cite{EFRB10} presented  the first non-blocking implementation
(which we will call \NBBST)
of a BST 
from single-word \CAS. \NBBST\ has several nice properties. Updates
operating on different parts of the tree do not interfere with one other
and 
\Find s 
never interfere with any other operation.
The code of \NBBST\ is modular and 
a detailed proof of correctness is provided in~\cite{techrept}.

In this paper, we build upon \NBBST\ to get a persistent version of it, called \NBBSTRQ. 
In a {\em persistent} data structure, old versions of the data structure are preserved when it is modified,
so that one can access any old version.
We achieve persistence on top of \NBBST\ by applying a relatively simple technique
which fully respects the modularity and simplicity of \NBBST's design.


In a concurrent setting, a major motivation for providing data structure persistence
is that it facilitates the implementation, in a wait-free way~\cite{H91}, of advanced operations
(such as range queries) on top of the data structure.
We exploit persistence in \NBBSTRQ\ to provide the first {\em wait-free} implementation 
of \Scan\ on  top of tree data structures, using single-word \CAS. 
\Scan($a$, $b$) returns a set containing all keys 
in the implemented set that are between the given keys $a$ and $b$.
\NBBSTRQ\ also provides {\em non-blocking}
(also known as {\em lock-free}~\cite{H91}) implementations of \Insert, \Delete, and \Find. 

\NBBSTRQ\ is {\em linearizable}~\cite{HW90}, uses single-word \CAS,
and tolerates any number of crash failures.  
As in \NBBST, 
updates in \NBBSTRQ\  on different parts of the tree 
are executed in parallel without  interfering with one another. 
A \Find\ simply follows tree edges from the root to a leaf and it may have 
to help an update operation only if the update is taking place at the parent or grandparent of the leaf
that the search arrives at. 
Thus, \Find\ employs a lightweight helping mechanism. 
Similarly, \Scan\ helps only those operations that are in progress
on the nodes that it traverses.
\Scan\ may print keys (or perform some processing of the nodes, e.g., counting them) 
as it traverses the tree, thus avoiding any space overhead.
\NBBSTRQ\ does not require knowledge of the number of processes in the system,
and therefore it works in a dynamic system where the set of participating processes changes.  

The code of \NBBSTRQ\ is as modular as that of \NBBST, making it fairly easy to understand.
However, designing a linearizable implementation of \Scan\ required solving 
several synchronization problems between \Scan s and concurrent update operations 
on the same part of the tree, 
so that a \Scan\ sees all the successful update operations  linearized before it 
but not those  linearized after it. 
Specifically, we had to (a) apply a mechanism based on sequence numbers set by \Scan s,
to split the execution into phases and assign each operation to a distinct phase, 
(b) design a scheme for linearizing operations that is completely different from that of of \NBBST\
by taking into consideration the phase to which each operation belongs, 
(c) ensure some additional necessary synchronization between \Scan s and updates,
and (d) use a more elaborate helping scheme. The proof of correctness 
borrows from that of \NBBST. However, due to the mentioned complications, 
many parts of it are more intricate.
The proof that \Scan s work correctly is completely novel.
\vspace*{-2mm}

\section{Related Work}

Our implementation is based on \NBBST, the binary search tree implementation
proposed in~\cite{EFRB10}. 
Brown {\em et al.}~\cite{BER13} generalized the techniques in~\cite{EFRB10}
to get the primitives \LLX, \SCX\ and \VLX\ which are generalizations of load-link, 
store-conditional and validate. These primitives can be used 
to simplify the non-blocking implementation of updates 
in every data structure based on a down tree (see~\cite{BER14,HL16} for examples). 
Unfortunately, our technique for supporting range queries
cannot directly be implemented using
\LLX\ and \SCX: the functionality hidden inside \LLX\ must be split
in two parts between which some 
synchronization is necessary 
to coordinate \Scan s with updates.
%
The work in~\cite{EFRB10} has also been generalized in~\cite{Sha13}
to get a non-blocking implementation of a Patricia trie.
None of these implementations of non-blocking search trees 
supports range queries.

%

Prokopec {\em et al.}~\cite{PBBO12} presented 
a non-blocking implementation of a concurrent hash trie which supports
a \FScan\ operation that provides a consistent
snapshot of the {\em entire} data structure.
Their algorithm 
uses indirection nodes (i-nodes)~\cite{V95} that double the height of the tree. 
To implement \FScan, the algorithm 
provides a persistent implementation of the trie
in which updates
may have to copy the entire path of nodes they
traverse to synchronize with concurrent \FScan s. 
Moreover, the algorithm causes a lot of contention on the root node.
The algorithm could be adjusted to support \Scan. 
However, every \Scan\ would cause updates taking place anywhere in the tree 
to copy all the nodes they visit, even if they are not in the part of the tree being scanned.

Petrank and Timnat~\cite{PT13} gave a technique
(based on 
\cite{J05})
to implement \FScan\ on top of non-blocking set data structures 
such as linked lists and skip lists.
Concurrent \FScan s  share a {\em snap collector} object 
in which they record information about the nodes they traverse.
To ensure that a \FScan\ appropriately synchronizes with 
updates, 
processes executing updates or \Find s must also record information about the operations they perform
(or those executed by other processes they encounter)
in the snap collector object. 
%
Although the 
snap collector object's primitive operations 
is wait-free, the following example shows that
the implementation of \FScan\ using those primitives is non-blocking but not wait-free.
Assume that the algorithm is applied on top of the non-blocking sorted linked list implementation 
presented by Harris~\cite{H01}.
A \FScan\ must traverse the list, and this traversal may never complete if concurrent updates continue to add more elements to the end of the list faster than the \FScan\ can traverse them.
In this case, the lists maintained in the snap collector will grow infinitely long.
In case $n$ is known, updates on different parts of the data structure
do not interfere with one another and have been designed to be fast.
However, \FScan\ is rather costly in terms of both time and space. 
\ignore{
\here{They claim their work generalizes to other data structures.  
Check Timnat's thesis to see if it contains the details. }
}
Chatterjee~\cite{Cha17} generalizes the algorithm of Petrank and Timnat to get
a non-blocking implementation of \Scan\ using partial snapshots~\cite{AGR08}. 
In a different direction, work in~\cite{ALRS17,Ros} characterizes when 
implementing the technique of~\cite{PT13}
on top of non-blocking data structures is actually possible.

Brown {\em et al.}~\cite{BA12} presented an implementation of a $k$-ary search tree 
supporting \Scan\ in an {\em obstruction-free} way~\cite{HLM03}. 
Avni {\em et al.}~\cite{ASS13} presented 
a skip list implementation which supports \Scan. It
can be either lock-free or be built on top of a transactional memory system,
so its progress guarantees are weaker than wait-freedom.
Bronson {\em et al.}~\cite{BCCO10} presented 
a {\em blocking} implementation of a relaxed-balance AVL tree
which provides support for \FScan.

Some papers present {\em wait-free} implementations of \FScan\ (or \Scan)
on data structures other than trees or in different settings. 
Nikolakopoulos {\em et al.}~\cite{NGPT15b,NGPT15a} 
gave a set of consistency definitions for \FScan\ 
and presented
\FScan\ algorithms for the lock-free concurrent queue
in~\cite{MS96}  that
ensure different consistency and progress guarantees.
Fatourou {\em et al.}~\cite{FNP17} presented a wait-free implementation 
of \FScan\ on top of the non-blocking deque implementation of~\cite{M03}.
Kanellou and Kallimanis~\cite{KK15} 
introduced a new graph model and provided a wait-free implementation
of a node-static graph 
which supports partial traversals 
in addition to edge insertions, removals, and weight updates.
Spiegelman {\em et al.}~\cite{SK16} presented 
two memory models and provided
wait-free dynamic atomic snapshot algorithms
for both.

\ignore{
In \NBBSTRQ, a process executing \Scan\ could report the
items it traverses on the fly without incurring any 
space overhead. Moreover, \Scan\ is more general than \FScan\ and
its implementation is {\em wait-free},
knowledge of $n$ is not needed, and updates operating
in different parts of the list never interfere with one another.
}

\ignore{
Maybe some citation for the idea of persistent data structures.  The usual way of doing persistence (creating a new root) wouldn't give much concurrency, so we use a different technique.  (Look for places where this kind of multi-version idea has been used before in distributed computing.)
}

\ignore{
\cite{BA12}:  quite a simple idea:  before removing a leaf, write a dirty bit in it.  Range query stores list of leaves whose interval intersects range.  Then go back and check dirty bit of all those leaves.  If none are set, then done; else retry.
}

\section{Overview of the BST Implementation and Preliminaries}
\label{sec:bst}

We provide a brief description of \NBBST\ (following the presentation in~\cite{EFRB10})
and some preliminaries.

\NBBST\ implements Binary Search Trees (BST) that are {\em leaf-oriented}, i.e.,
%
all keys 
are stored in the leaves of the tree.
The tree is full and maintains the {\em binary search tree property}:
for every node $v$ in the tree, 
the key of $v$ is larger than the key of every node in $v$'s left subtree 
and smaller than or equal to the key of every node in $v$'s right subtree. 
The keys of the Internal nodes are used solely for routing to the appropriate leaf
during search.
A leaf (internal) node is represented by an object of type Leaf (Internal, respectively);
we say that Leaf and Internal nodes are of type Node  (see Figure~\ref{code1}).

To insert a key $k$ in a leaf-oriented tree, 
a search for $k$ is first performed. 
Let $\ell$ and $p$ be the leaf that this search arrives at and its parent. 
If $\ell$ does not contain $k$, then a subtree consisting of an internal node and
two leaf nodes is created. The leaves contain $k$ and the key of $\ell$
(with the smaller key in the left leaf). The internal node
contains the bigger of these two keys. 
The child pointer of $p$ which was pointing to $\ell$ is changed to point to 
the root of this subtree. 
Similarly, for a \Delete($k$), let $\ell$, $p$ and $gp$
be the leaf node that the search \Delete\ performs arrives at, its parent, 
and its grandparent. If the key of 
$\ell$ is $k$, then the child pointer of $gp$ which was pointing to $p$
is changed to point to the sibling of $\ell$. 
By performing the updates in this way, 
the properties of the tree
are maintained.

An implementation is {\em linearizable} if, in every execution $\alpha$,
each operation that completes in $\alpha$ (and some that do not)
can be assigned a {\em linearization point} between the starting and finishing time of its execution so that the return values of those operations 
are the same in $\alpha$ as if the operations were executed
sequentially in the order specified by their linearization points. 

To ensure linearizability, \NBBST\ applies a technique that flags and marks 
nodes. A node is flagged before any of its child pointers changes. A node is permanently marked before it is removed.
To mark and flag nodes,
\NBBST\ uses \CAS. \CAS($O, u, v$) changes the value of object $O$ to $v$
if its current value is equal to $u$, otherwise the \CAS\ fails and no change is applied on $O$. 
In either case, the value that $O$ had before the
execution of \CAS\ is returned.  
\remove{Each node has a field, called $type$; 
this field takes the following values: \clean\ when the node is neither flagged
nor marked, \mk\ when the node is marked, and \IFlag\ or \DFlag\ when the node
is flagged by an insert or a delete operation, respectively. }

\NBBST\ provides a routine, \Search($k$),
to search the data structure for key $k$. \Search\ returns 
pointers to the leaf node at which the \Search\ arrives,
to its parent, and to its grandparent. 
\Find($k$) executes \Search($k$) and checks whether the returned leaf contains 
the key $k$. 
\Insert($k$) executes \Search($k$) to get a leaf $\ell$ and 
its parent $p$. It then performs a {\em flag} \CAS, to flag $p$, 
then a {\em child} \CAS\ to change the appropriate child pointer of $p$ 
to point to the root of the newly created subtree of three nodes, 
and finally an {\em unflag} \CAS\ to unflag $p$.
If it fails to flag $p$, it restarts without executing
the other two \CAS\ steps. 
%
Similarly, a \Delete($k$) calls \Search\ to get a leaf $\ell$, 
its parent $p$, and its grandparent $gp$. It first executes a flag \CAS\ trying to flag $gp$.
If this fails, it restarts. If the flagging succeeds, it executes a {\em mark} \CAS\ to mark $p$.
If this fails, it unflags $gp$ and restarts. 
Otherwise, it executes a child \CAS\ to change the apropriate child pointer of $gp$
to point from $p$ to the sibling of $\ell$,
it unflags $p$ and returns. 
Both \Insert\ and \Delete\ operations execute the body of a while loop repeatedly until they succeed.
The execution of an iteration of the while loop is called {\em attempt}.
%

Processes may fail by {\em crashing}.
An implementation is {\em non-blocking} if in every infinite execution,
infinitely many operations are completed.
\NBBST\ is non-blocking: Each process $p$ that flags
or marks a node stores in it a pointer to an Info object, which contains information about 
the operation $op$ it performs (see Figure~\ref{code1}). 
This information includes the old and new values
that should be used by the \CAS\ steps that $p$ will perform to complete
the execution of $op$.
Other processes that apply operations on the same part of the data structure
can help this operation complete and unflag the node. 
Once they do so, they are able
to retry their own operations. 
Helping is necessary only if an update operation
wants to flag or mark a node already flagged or marked by another process.

\ignore{
To avoid handling special cases, 
\NBBST\ assumes that the implemented set always contains 
two special values, namely $\infty_1, \infty_2$,
where $\infty_1 < \infty_2$ and every key other than these values in the universe
is less than $\infty_1$. These two keys are never removed from the tree,
so the tree always contains at least three nodes (see Figure~\ref{+++}).

To linearize \Insert, \Delete, and \Find\ operations, \NBBST\ assigns
linearization points to instances of \Search. Specifically, it is proved 
in~\cite{EFRB10} that for each node 
visited by an instance of \Search, there has been a point in time after the beginning
of \Search\ that this node was in the tree (although it may have been removed from the tree
by the time that the \Search\ visits it). \Search\ is linearized at the latest
point in time that the leaf $\ell$ it returns was in the tree
(and no later than the time that \Search\ visits this leaf). 
A \Find\ operation is linearized at the same time that the \Search\ it executes is linearized.
Each \Insert() operation that sucessfully flags a node 
is linearized at the point that the successful child \CAS\ for this \Insert\ is performed. 
Similarly, each \Delete\ operation that sucessfully marks a node 
is linearized at the point that the successful child \CAS\ for this \Delete\ is performed. 
(It is proved in~\cite{EFRB10} that there is exactly one successful child \CAS\
for every such update.)
}

\section{A Persistent Binary Search Tree Supporting Range Queries}
\label{sec:alg}

\begin{figure}[t]
\begin{center}
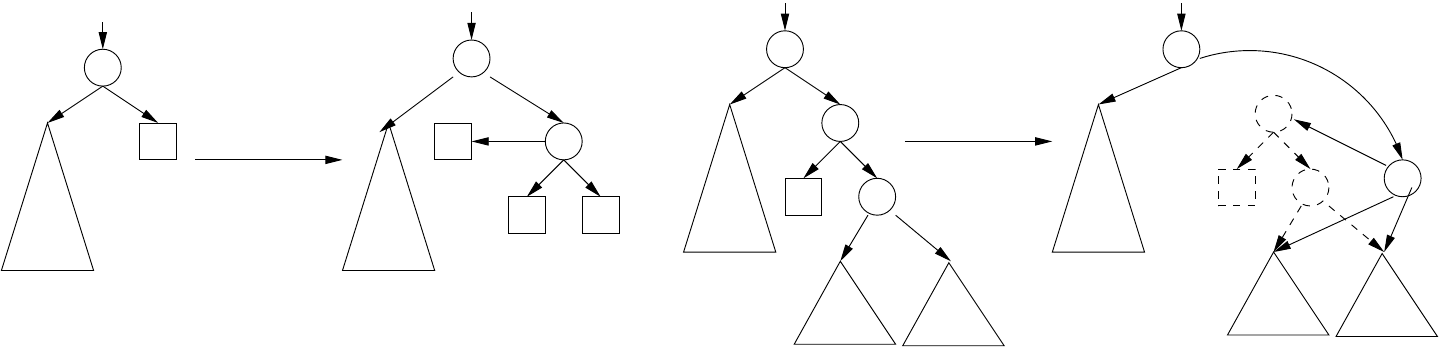
\end{center}
\capspace
\caption{\label{ins-example}Examples of \Insert\ and \Delete.}
\end{figure}

We modify \NBBST\ to get \NBBSTRQ, a BST implementation
that supports \Scan, in addition to \Insert, \Delete, and \Find.

\subsection{Overview}

%
In a concurrent environment, care must be taken to synchronize 
\Scan s with updates since
as a \Scan\ traverses the tree, it may see an update $op$ by a process $p$
but it may miss an update that finishes before $op$ starts, 
and was applied on the part of the tree that has already been visited by the \Scan\
(thus violating linearizability).

To avoid such situations, \NBBSTRQ\ implements
a persistent version of the leaf-oriented tree,
thus allowing a \Scan\ to reconstruct previous versions of it.
To achieve this, \NBBSTRQ\ stores in each node an additional pointer, called $prev$.
Whenever the child pointer of a node $v$ changes from a node $u$
to a node $u'$, the $prev$ pointer of  $u'$ points to $u$.  
(Figure~\ref{ins-example} illustrates an example.) 

\NBBSTRQ\ maintains a shared integer, $Counter$, which is incremented
each time a \Scan\ takes place.
Each operation has a {\em sequence number} associated with it. 
Each \Scan\  starts its execution by reading $Counter$ and uses the 
value read as its sequence number. 
Each other operation $op$ reads $Counter$ at the beginning of each of its attempts.
The sequence number of $op$ is the sequence number read in its last attempt.
A successful update operation
records its sequence number in the Info object 
it creates during its last attempt.
Intuitively, each \Scan\ initiates a new execution phase whenever it increments $Counter$. 
For each $i \geq 0$, 
phase $i$ is the period during which $Counter$ has the value $i$.
We say that all operations with sequence number $i$ 
belong to phase $i$.

Each tree node 
has a {\em sequence number} which is the sequence number of the 
operation that created it. In this way, a \Scan\ 
may figure out which nodes have been inserted  or 
 deleted  by updates that belong to later phases. 
For any Internal node $v$ whose sequence number is at most $i$, we define the {\it version-$i$ left (or right) child of $v$} to be the node that is reached by following the
left (or right) child pointer of $v$ and then following its $prev$ pointers until reaching the first node whose $seq$ field
is less than or equal to $i$.  
(We prove that such a node exists.) 
For every configuration $C$, we define graph $D_{i}(C)$ as follows.
The nodes of $D_{i}(C)$ is the set of all existing nodes in $C$ and the edges go from nodes to their version-$i$ children;
$T_i(C)$ is the subgraph of $D_i(C)$ containing those nodes that are reachable from the root node in $D_i(C)$. 
We prove that
$T_i(C)$ is a binary search tree.

We linearize every \func{Scan} operation with sequence number $i$ 
at the end of phase $i$,
with ties broken in an arbitrary way. 
Moreover, we linearize all \Insert, \Delete\ and \Find\ operations that belong to phase $i$
during phase $i$.
To ensure linearizability, \NBBSTRQ\ should guarantee that 
a \Scan\ with sequence number $i$  ignores all changes 
performed by successful update operations that belong to phases with sequence numbers bigger than $i$.
To ensure this, 
each operation with sequence number $i$ ignores
those nodes of the tree that have sequence numbers bigger than $i$ 
by moving from a node to its appropriate version-$i$ child.
Thus, each operation with sequence number $i$ always operates on $T_i$.

\begin{figure}[bt]
{\footnotesize
\begin{minipage}{0.55\textwidth}
\begin{code}
\firstline
type Update \{ \hspace*{14mm} \com stored in one CAS word\nl
\n  $\{\flag,\mk\}$ $type$ \nl
    Info\ *{\it info}\nl
\p
\}
\nl
type Info\ \{  \nl
\n  $\{\bot, \try,\commit, \abort\}$ $state$ \nl 
    Internal *$nodes$[] 	\tabcom nodes to be frozen \nl 
    Update $oldUpdate$[]	\tabcom old values for freeze CAS steps \nl
    Internal *$mark$[]		\tabcom nodes to be marked \nl
    Internal *$par$ 		\tabcom node whose child will change \nl
    Node *$oldChild$		\tabcom old value for child \CAS\nl  
    Node *$newChild$		\tabcom new value for the child \CAS\nl 
    int $seq$ 			\tabcom sequence number \nl
\p\}
\end{code}
\end{minipage}
\hfill
\begin{minipage}{0.45\textwidth}
\begin{code}
\firstline
type Internal \{ \hspace*{13.25mm} \com subtype of Node\nl
\n $\Key \cup \{\infty_1,\infty_2\}$ $key$\nl
   Update $update$\nl
   Node *\lft, *$right$\nl
   Node *$prev$\nl
   int seq\nl
\p\}
\nl
type Leaf \{ \hspace*{18.3mm} \com subtype of Node\nl
\n $\Key \cup \{\infty_1,\infty_2\}$  $key$\nl
   Update $update$\nl
   Node *$prev$\nl
   int seq\nl
\p\}
\end{code}
\end{minipage}

\vspace*{-2mm}

\begin{code}
\firstline
\com Initialization:\nl
shared counter $Counter$ := 0\nl
shared Info *$Dummy$ := pointer to a new Info object whose $state$ field is \abort, and whose other fields are \NULL \nl
shared Internal *$Root$ := pointer to new Internal node with $key$ field $\infty_2$, $update$ field $\langle \flag,Dummy\rangle$,\ul
\n  $prev$ field  \NULL, $seq$ field $0$, 
and its $left$ and $right$ fields pointing to new Leaf nodes whose $prev$ fields\ul 
 are \NULL, $seq$ fields are $0$, and keys $\infty_1$ and $\infty_2$, respectively 
\p 
\end{code}
}
\vspace*{-6mm}
\caption{\label{code1}Type definitions and initialization.}
\end{figure}

\begin{figure*}
{\footnotesize
\begin{code}
\firstline
\func{Search}($\Key\ k$, int $seq$): $\langle \mbox{Internal*}, \mbox{Internal*}, \mbox{Leaf*}\rangle$  \{\nl
\n     \com Precondition:  $seq\geq 0$\nl
	   Internal *$gp$, *$p$\nl
       Node *$l:= Root$\label{search-initialize}\nl

       while $l$ points to an internal node \{ \label{search-while}\nl
\n         $gp := p$ \tabtabcom Remember parent of $p$ \label{search-advance-gp}\nl
           $p := l$ \tabtabcom Remember parent of $l$ \label{search-advance-p}\nl
           $l = \func{ReadChild}(p, k < p \rightarrow key, seq)$ \tabtabcom Go to appropriate version-$seq$ child of $p$\label{search-advance-l}\nl 
       \p \} \nl
       return $\langle gp, p, l \rangle$\label{search-return} \nl
\p 
\}\bl
\nl
\func{ReadChild}($Internal * p, \mbox{Boolean } left, \mbox{int }seq$): $\mbox{Node*}$ \{ \nl
\n  \com Precondition:  $p$ is non-\NULL\ and $p\rightarrow seq \leq seq$\nl
    if $left$ then  $l:= p\rightarrow \lft$ else $l:=p \rightarrow right$ \label{read-child}\tabtabcom Move down to appropriate child\nl
    while ($l \rightarrow seq > seq$)  $l := l \rightarrow prev$ \label{rc-while} \label{read-prev}\nl
return $l$; \nl
\p
\} \ul
\nl
\func{ValidateLink}($\mbox{Internal *} parent, \mbox{Internal *} child, \mbox{Boolean }left$): $\langle \mbox{Boolean, Update}\rangle$ \{ \nl
\n \com Preconditions:  $parent$ and $child$ are non-\NULL\ \nl
   Update $up$
\nl
   $up := parent \rightarrow update$ \label{val-read-pupdate}\nl
   if $\func{Frozen}(up)$ then \label{val-check-frozen}\{\nl
\n      \func{Help}($up.info$) \label{val-help}\nl
        return $\langle \FALSE, \bot \rangle$\nl
\p \}\nl
   if ($left$ and $child \neq parent\rightarrow left$) or ($\neg left$ and $child \neq parent\rightarrow right$) then return $\langle \FALSE, \bot\rangle$ \label{val-read-child}\nl
   else return $\langle \TRUE, up\rangle$\nl
\p \} \ul
\nl
\func{ValidateLeaf}(Internal *$gp$, Internal *$p$, Leaf *$l$, Key $k$) :  $\langle\mbox{Boolean, Update, Update}\rangle$ \{\nl
\n   \com Preconditions:  $p$ and $l$ are non-$\bot$ and if $p\neq Root$ then $gp$ is non-$\bot$\nl
     Update $pupdate, gpupdate := \bot$\nl
     Boolean $validated$     \nl
     $\langle validated, pupdate\rangle := \func{ValidateLink}(p,l,k<p\rightarrow key)$ \label{validate-leaf-p}\nl
     if $validated$ and $p\neq Root$ then $\langle validated, gpupdate\rangle := \func{ValidateLink}(gp,p,k<gp\rightarrow key)$\label{validate-gp-p}\nl
     $validated := validated \mbox{ and } p\rightarrow update = pupdate \mbox{ and }(p = Root \mbox{ or }gp\rightarrow update = gpupdate)$ \label{validate-reread}\nl
     return $\langle validated, gpupdate, pupdate\rangle$\nl
\p \}\ul
\nl
\func{Find}($\Key\ k$): Leaf* \{ \nl
\n   Internal * $gp,p$\nl 
     Leaf *$l$\nl
     Boolean $validated$\ul
\nl
     while \TRUE\ \{ \label{find-while}\nl
\n     $seq := Counter $ \label{find-getseq}\nl
	   $\langle -,p,l\rangle := \func{Search}(k, seq)$ \label{find-search}\nl
       $\langle validated,-, -\rangle := \func{ValidateLeaf}(gp,p, l, k)$ \label{find-validate-leaf}\nl
       if $validated$ then \{ \nl
\n          if $l\rightarrow key = k$ then return $l$\nl
            else return \NULL\nl
\p \} \nl
\p \} \nl
\p
\}
\ul
\nl
\func{CAS-Child}(Internal *$parent$, Node *$old$, Node *$new$) \{\ul
\n  \com {\it Precondition}:  $parent$ points to an Internal node and $new$ points to a Node (\ie, neither is \NULL) and $new\rightarrow prev=old$\ul
    \com This routine tries to change one of the child fields of the node that $parent$ points to from $old$ to $new$.
\nl
    if $new \rightarrow key < parent\rightarrow key$ then\label{which-child}\nl
\n       \CAS$(parent\rightarrow \lft, old, new)$\tabtabcom{\bf child CAS}\label{childCAS1}\nl
\p  else\nl
\n       \CAS$(parent\rightarrow right, old, new)$\tabtabcom{\bf child CAS}\label{childCAS2}\nl
\p\p 
\} \ul
\end{code}
}
\vspace*{-5mm}
\caption{\label{code2}Pseudocode for \func{Search}, \func{Find} and some helper routines.}
\end{figure*}

\begin{figure*}
{\footnotesize
\begin{code}
\firstline
\func{Frozen}(Update $up$): Boolean \{ \nl
\n return (($up.type = \flag$ and  $up.info \rightarrow state \in \{ \bot, \try\})$ or \ul
\n\n $(up.type = \mk$ and $up.info \rightarrow state \in \{ \bot, \try, \commit\}$)) \label{frozen}\nl
\p\p
\p 
\}\ul\nl
\func{Execute} (Internal *$nodes$[], Update $oldUpdate$[], Internal *$mark$[], Internal *$par$, \ul
\n \n \n \n \n Node *$oldChild$, Node *$newChild$, int $seq$): Boolean \{ \nl
\p \p \p \p \p
\n   \com Preconditions:  (a) Elements of $nodes$ are non-\NULL, (b) $mark$ is a subset of $nodes$, (c) $par$ is an element of $nodes$,\nl
\n \n               \com  (d) $oldChild$ and $newChild$ are distinct and non-\NULL, (e) $oldChild$ is an element of $mark$,  \nl
					\com (f) $newChild\rightarrow prev = oldChild$, and (g) if $par=Root$ then $newChild\rightarrow key$ is infinite.\nl
\p\p for $i := 1$ to length of $oldUpdate$ \{ \label{execute-for}\nl
\n       if $\func{Frozen}(oldUpdate[i])$  then \{ \label{execute-check-frozen}\nl
\n           if $oldUpdate[i].info\rightarrow state\in\{\bot,\try\}$ then \func{Help}($oldUpdate[i].info$) \label{execute-help-others}\nl 
             return \FALSE \label{execute-false}\nl
\p       \} \nl
\p \} \nl
    $infp :=$ pointer to a new Info\ \record\  containing $\langle \bot, nodes, oldUpdate, mark, par, oldChild, newChild, seq\rangle$\label{create-info}\nl
    if $\CAS(nodes[1] \rightarrow update, oldUpdate[1], \langle \flag, infp \rangle)$ then  \tabtabcom {\bf freeze CAS} \label{flagCAS1}\nl
\n        return \func{Help}($infp$)  \label{execute-help-self}\nl
\p  else return \FALSE \nl
\p\} \ul
\nl
\func{Help}(Info *$infp$): boolean  \{ \nl
\n   \com Precondition:  $infp$ is non-\NULL\ and does not point to the Dummy Info object\nl
     int $i := 2$ \nl
     boolean $continue$ \bl \nl

     if $Counter \neq infp \rightarrow seq$ then\label{help-handshaking}\nl
\n       \CAS($infp \rightarrow state$, $\bot$, \abort) \tabtabcom{\bf abort CAS}\label{abortCAS}\nl
\p   else
       \CAS($infp \rightarrow state$, $\bot$, \try)\tabtabcom{\bf try CAS}\label{tryCAS}\nl  
 
     $continue := (infp\rightarrow state = \try)$ \label{help-check-try1}\nl
     while $continue$ and $i \leq$ length of $infp \rightarrow nodes$ do \label{help-while-begin}\{ \nl
\n        if $infp\rightarrow nodes[i]$ appears in $infp\rightarrow mark$ then \nl
\n            $\CAS(infp \rightarrow nodes[i]\rightarrow update, infp \rightarrow oldUpdate[i], \langle \mk, infp\rangle)$ \tabtabcom {\bf freeze \CAS} \label{markCAS}\nl
\p        else\
            $\CAS(infp \rightarrow nodes[i]\rightarrow update,  infp \rightarrow oldUpdate[i], \langle \flag, infp \rangle)$ \tabtabcom {\bf freeze \CAS} \label{flagCAS2}\nl
        $continue := (infp \rightarrow nodes[i]\rightarrow update.info = infp)$ \label{help-check-frozen} \nl
          $i := i + 1$ \nl
\p \} \label{help-while-end}\nl
    if $continue$ then \{ \label{help-check-continue}\nl
\n       $\func{CAS-Child}(infp \rightarrow par, infp \rightarrow oldChild, infp \rightarrow newChild)$\label{help-CAS-child}\nl
         $infp\rightarrow state := \commit$		 \tabtabcom {\bf commit write}\label{commitWrite} \nl
\p \} else if $infp\rightarrow state = \try$ then \label{help-check-try2}\nl
\n       $infp\rightarrow state := \abort$ 	\tabtabcom{\bf abort write} \label{abortWrite}	\nl
\p return ($infp\rightarrow state = \commit$) \label{help-return}\nl 
\p  
\} 
\ul
\nl
\Scan(int $a$, int $b$): Set \{ \nl
\n    $seq := Counter$  \label{scan-getseq}\nl
      $Inc(Counter)$ \label{scan-inc}\nl
      return $\func{ScanHelper}(Root, seq, a, b)$ \label{scan-return}\nl
\p \} \ul
\nl
\func{ScanHelper}(Node *$node$, int $seq$, int $a$, int $b$): Set \{ \nl
\n   \com Precondition:  $node$ points to a node with $node\rightarrow seq \leq seq$\nl
     Info * $infp$\bl\nl
     if $node$ points to a leaf then return $\{ node \rightarrow key\} \cap [a,b]$\label{scanhelper-leaf}\nl
     else \{\nl
\n       $infp := node\rightarrow update.info$ \label{scanhelper-read-info}\nl
         if $infp\rightarrow state\in\{\bot,\try\}$ then $\func{Help}(infp)$ \label{scanhelper-help}\nl
         if $a > node \rightarrow key$ then  return \func{ScanHelper}$(\func{ReadChild}(node, \FALSE, seq), a,b)$\label{scanhelper-recursive1}\nl
         else if $b < node \rightarrow key$ then return \func{ScanHelper}$(\func{ReadChild}(node, \TRUE, seq), a, b)$ \label{scanhelper-recursive2}\nl
         else return $\func{ScanHelper}(\func{ReadChild}(node, \FALSE, seq), a, b)$ $\cup$ \label{scanhelper-recursive3}\nl 
\n \n                $\func{ScanHelper}(\func{ReadChild}(node,\TRUE, seq), a, b)$ \label{scanhelper-recursive4}\nl
\p \p \p   \}  \nl
\p \} 
\end{code}
}
\vspace*{-5mm}
\caption{\label{code:ann-help}Pseudocode for \func{Execute}, \func{Help} and \func{Scan}.}
\end{figure*}

\begin{figure*}
{\footnotesize
\begin{code}
\firstline
\func{Insert}($\Key\ k$): boolean \{ \nl
\n Internal * $gp$, *$p$, *$newInternal$\nl 
   Leaf *$l$, *$newSibling$\nl 
   Leaf *$new$ \nl
   Update $pupdate$\nl
   Info *$infp$\nl
   Boolean $validated$\ul
\nl
   while \TRUE\ \{ \label{insert-while} \nl
   \n $seq := Counter $ \label{insert-getseq}\nl
      $\langle gp, p, l \rangle := \func{Search}(k,seq)$ \label{insert-search}\nl
      $\langle validated, -,pupdate \rangle := \func{ValidateLeaf}(gp,p,l,k)$ \label{insert-validate-leaf}\nl
      
      if $validated$ then \{\nl
\n       if $l \rightarrow key = k$ then return \FALSE \tabtabcom Cannot insert duplicate key\label{insert-false}\nl
         else \{ \nl
\n          $new  :=$ pointer to a new Leaf node whose $key$ field is $k$, its $seq$ field is equal to $seq$, and its $prev$ field is \NULL \label{create-new-leaf}\nl
            $newSibling :=$ pointer to a new Leaf whose key is $l\rightarrow key$, \label{create-leaf}\ul
\n                        its $prev$ field is equal to \NULL\ and its $seq$ field is equal to $seq$\label{create-sibling-leaf}\nl
\p          $newInternal :=$ pointer to a new Internal node with $key$ field $\max(k, l \rightarrow key)$,\label{create-internal}\ul      
\n                               $update$ field $\langle \flag, Dummy\rangle$, its $seq$ field equal to $seq$ and its $prev$ field equal to $l$, \ul
                           and with two child fields equal to $new$ and $newSibling$\ul 
          	           (the one with the smaller key is the left child),  \nl
\p             if \func{Execute}($[p, l], [pupdate, l\rightarrow update], [l], p, l, newInternal, seq$) then return \TRUE \label{insert-execute}\nl
\p      \}  \nl
\p    \}\nl
\p \}\nl
\p
\}\ul
\nl
\func{Delete}($\Key\ k$): boolean \{\nl
\n Internal *$gp$, *$p$\nl
   Leaf *$l$\nl
   Node *$sibling$, *$newnode$ \nl
   Update $pupdate, gpupdate, supdate$\nl
   Info *$infp$\nl
   Boolean $validated$\bl\nl

   while \TRUE\ \{ \label{delete-while}\nl
   \n $seq := Counter $ \label{delete-getseq}\nl
   $\langle gp, p, l\rangle := \func{Search}(k, seq)$\label{delete-search}\nl
   $\langle validated, gpupdate, pupdate\rangle := \func{ValidateLeaf}(gp,p,l,k)$ \label{delete-validate-leaf}\nl
   if $validated$ then \{\nl
\n       if $l\rightarrow key \neq k$ then return \FALSE \tabtabcom Key $k$ is not in the tree\label{delete-false}\nl
         $sibling$ := \func{ReadChild}($p, l \rightarrow key \geq p\rightarrow key, seq$) \label{read-sibling}\nl
         $\langle validated, -\rangle := \func{ValidateLink}(p, sibling, l\rightarrow key \geq p\rightarrow key)$ \label{delete-validate-p-sib}\nl
         if $validated$ then \{\nl
\n           $newNode :=$ pointer to a new copy of sibling with its $seq$ field set to $seq$ and its $prev$ pointer set to $p$ \label{copy-sibling}\nl
             if $sibling$ is Internal then \{\nl
\n               $\langle validated , supdate\rangle := \func{ValidateLink}(sibling, newNode\rightarrow left, \TRUE)$\label{validate-sib-nephew1}\nl
                 if $validated$ then $\langle validated, -\rangle := \func{ValidateLink}(sibling, newNode\rightarrow right, \FALSE)$\label{validate-sib-nephew2}\nl
\p           \} else $supdate = sibling\rightarrow update$ \label{read-sibling-update}\nl
             if $validated $ and \func{Execute}($[gp, p, l, sibling], [gpupdate, pupdate, l\rightarrow update, supdate],$   \label{delete-execute}\ul
  \hspace*{39mm}$[p, l, sibling], gp, p, newNode, seq$) then \nl 
\n                   return \TRUE \label{del-succeed} \nl
\p \p     \}\nl
\p        \}\nl
\p   \} \nl
\p \}
\end{code}
}
\caption{\label{code3}Pseudocode for \func{Insert} and \func{Delete}.}
\end{figure*}

\later{Terminology about frozen/flagged/marked is confusing.  Figure out good terminology and stick with it.}

To ensure linearizability, \NBBSTRQ\ should also ensure that each \Scan\  sees
all the successful updates that belong to phases smaller than or equal to $i$.
To achieve this, \NBBSTRQ\ employs a handshaking mechanism between each scanner 
and the updaters. It also uses a helping mechanism which is more elaborate than 
that of \NBBST. 

To describe the handshaking mechanism in more detail,
consider any update operation $op$ initiated by process $p$. 
No process can be aware of $op$ before $p$
performs a successful flag \CAS\ for $op$. 
Assume that $p$ flags node $v$ for $op$
in an attempt $att$ with sequence number $i$. 
To ensure that no \Scan\ with sequence number $i$ will miss $op$, 
$p$ checks whether $Counter$ still has the value $i$ after the flag \CAS\ has occurred. 
We call this check the {\em handshaking check} of $att$. 
If the handshaking check succeeds, it is guaranteed 
that no \Scan\ has begun its traversal 
between the time that $p$ reads $Counter$ at the beginning of the execution of $att$ and the time the handshaking check
of $att$ is executed.
Note that any future \Scan\ with sequence number $i$ that traverses $v$ while $att$ is still in progress, 
will see that $v$ is flagged and find out the required information to complete $op$
in its Info object. In \NBBSTRQ, the \Scan\ helps $op$ complete
before it continues its traversal. 

However, if the handshaking check fails, 
$p$ does not know whether any \Scan\ that incremented $Counter$ to a value greater than $i$
has already traversed the part of the tree that $op$ is trying to update, 
and has missed this update. 
At least one of these \Scan s 
will have sequence number equal to $i$. Thus, if $op$ succeeds, 
linearizability could be violated.
To avoid this problem, 
$p$ pro-actively aborts its attempt of $op$ if the handshaking check fails, 
and then it initiates a new attempt for $op$ (which will have a sequence number bigger than $i$).
This abort mechanism is implemented as follows.
The Info object 
has a field, called $status$, which takes 
values from the set $\{\bot, \try, \commit, \abort \}$ (initially $\bot$). 
Each attempt creates an Info object. 
To abort the execution of an attempt, $p$ changes the $status$
field of its Info object to $\abort$.
Once an attempt is aborted, the value of the $status$ field
of its Info object remains \abort\ forever.
If the handshaking check succeeds, then $p$ changes the 
$status$ field of the Info object of $att$ to \try\ 
and tries to execute the remaining steps of this attempt.
If $op$ completes successfully, it changes
the $status$ field of the Info object to \commit.
Info objects whose $status$ field is equal to $\bot$ or \try\ belong to
update operations that are still {\em in progress}.

We now describe the linearization points in more detail.
If an attempt of an \Insert\ or \Delete\ ultimately succeeds in updating a child pointer of the tree to make the update take effect,
we linearize the operation at the time that attempt first flags a node:  this is when 
the update first becomes visible to other processes.
(This scheme differs from the original \NBBST, where updates are linearized at the time they actually change
a child pointer in the tree.)
Because of handshaking, this linearization point is guaranteed to be before the end
of the phase to which the operation belongs.

When a \Find\ operation completes a traversal of a branch of the tree to a leaf,
it checks whether an update has already removed the leaf or is in progress and could 
later remove that leaf from the tree.
If so, the \Find\ helps the update complete and retries.  
Otherwise, the \Find\ terminates and is linearized at the time when the leaf is in the tree and has
no pending update that might remove it later.
(As in the original \NBBST, the traversal of the branch may pass through nodes
that are no longer in the tree, but so long as it ends up at a leaf that is still present in the current tree
we prove that it ends up at the correct leaf of the current tree.)
An \func{Insert}($k$) that finds key $k$ is already in the tree, and a \func{Delete}($k$) that 
discovers that $k$ is not in the tree are linearized similarly to \Find\ operations.

\ignore{
To make the assignment of linearization points to \Insert, \Delete\ and \Find\ operations 
with the same sequence number $i$ more concrete, 
we first define a \emph{tentative linearization point} for each of these operations.
The assignment of tentative linearization points to \Find s and update operations
is done as in \NBBST. 
Specifically, we prove that there is a unique successful child \CAS\ for each successful update.
The tentative linearization point of an update 
that has a successful child \CAS\ is that successful child \CAS. 
We also prove that any node visited by a \func{Search}($k$, $i$) 
was on the search path for $k$ in $T_i(C)$ in some configuration $C$ during the \func{Search}.
The tentative linearization point of each \func{Find} that terminates or update that returns \FALSE\ 
is at the latest point (during its final \Search)
where the leaf that this \func{Search} ends at is in $T_i$. 
We linearize all terminated operations, 
plus all update operations that have a successful child \CAS\ 
in the order of their associated sequence numbers.  
Among operations with the same sequence number,
we first order all operations except \func{Scans} according to their tentative linearization point, 
and then put all \func{Scans} after them.
}

The helping mechanism employed by \func{Find} operations ensures that the \func{Find}
will see an update that has been linearized (when it flags a node) before the \func{Find} but has not yet
swung a child pointer to update the shape of the tree.
But it is also crucial for synchronizing with \Scan\ operations, for the following reason.
Assume that a process $p_1$ initiates an \Insert($1$).
It reads $0$ in $Counter$ and successfully performs its flag \CAS.
Then, a \Scan\ is initiated by a process $p_2$ 
and changes the value of $Counter$ from $0$ to $1$.
Finally, a \Find(1)\ is initiated by a process $p_3$
and reads $1$ in $Counter$. 
\Find($1$) and \Insert($1$) will arrive at the same leaf node $\ell$
(because \Insert($1$) has not performed 
its child \CAS\ by the time \Find\ reaches the leaf). 
If \Find($1$)\ ignores the flag that exists on the parent node of $\ell$ 
and does not help \Insert($1$) to complete, 
it will return \FALSE. 
If \Insert($1$) now continues its execution, 
it will complete successfully,
and given that it has sequence number $0$, it will be linearized before \Find($1$)
which has sequnce number $1$. That would violate linearizability. 

\subsection{Detailed Implementation}

A \Scan($a,b$) first determines 
its sequence number $seq$ (line~\ref{scan-getseq})
and then increments $Counter$ to 
start a new phase (line~\ref{scan-inc}). 
To traverse the appropriate part of the tree, it calls \ScanHelper($Root, seq, a,b$) (line~\ref{scan-return}).  
\ScanHelper\  
starts from the root and recursively calls itself 
on the version-$seq$ left child of the current node $v$ 
if $a$ is greater than $v$'s key, or on $v$'s version-$seq$ right child 
if $b$ is smaller than $v$'s key, 
or on both version-$seq$ children if $v$'s key is between $a$ and $b$ 
(lines~\ref{scanhelper-leaf}--\ref{scanhelper-recursive4}). 
Whenever it visits a node where an update is in progress,
it helps the update to complete (line~\ref{scanhelper-help}). 
\ReadChild\ is used to obtain $v$'s appropriate version-$seq$ child.

\Search($k,seq$) traverses a branch of $T_{seq}$ from the root to a leaf node
(lines~\ref{search-while}--\ref{search-advance-l}).
\Find\ gets a sequence number $seq$
(line~\ref{find-getseq}) and calls \Search($k$, $seq$) (line~\ref{find-search})
to  traverse the BST to a leaf $l$. 
Next, it calls \ValidateLeaf\  
to ensure that 
there is no update that has removed $l$ or has flagged $l$'s parent $p$ or grandparent $gp$
for an update that could remove $l$ from the tree.
If the validation succeeds, the \func{Find} is linearized at line \ref{validate-reread}.
If it finds an update in progress, the \func{Find} helps complete it at line \ref{val-help}. 
If the validation is not successful, 
\Find\ retries.

An \Insert($k$) performs repeated attempts.  Each attempt 
 first determines a sequence number $seq$, and 
calls \Search($k$, $seq$) (line~\ref{insert-search}) 
to traverse to the appropriate leaf $l$ in $T_{seq}$.
It then calls \ValidateLeaf, just as \Find\ does.
If the validation  
is successful 
and $k$ is not already in the tree (line~\ref{insert-false}), 
a subtree of three nodes is created (lines~\ref{create-new-leaf}--\ref{create-internal}).
\Execute\ (line~\ref{insert-execute}) performs the remaining actions of the \Insert, in a way that is similar to the \func{Insert} of \NBBST.

In a way similar to \Insert($k$), 
a \Delete($k$) performs repeated attempts (line~\ref{delete-while}).
Each attempt
determines 
its sequence number $seq$ (line~\ref{delete-getseq}) and
calls \Search($k$, $seq$) (line~\ref{delete-search}) 
to get the leaf $\ell$, its parent $p$ and grandparent $gp$. 
Next, it validates the leaf (as in \func{Find}).
If successful,
it finds the sibling of $\ell$ (lines~\ref{read-sibling}--\ref{read-sibling-update}) 
and calls \Execute\ (line~\ref{delete-execute})
to perform the remaining actions.
We remark that, in contrast to what happens in \NBBST\ which changes
the appropriate child pointer of $gp$ to point to the sibling of $\ell$, 
\NBBSTRQ\ creates a new node where it copies the sibling of $\ell$
and changes the appropriate child pointer of $gp$ to point to this new copy.
This is necessary to avoid creating cycles consisting of $prev$ and $child$ pointers, which could cause 
infinite loops during \Search.
\ignore{
The reason for this is the following. 
Suppose a node with key $7$ has a left child with key $5$ and sequence number $1$.
The node with key $3$ has two children with keys $1$ and $5$ and sequence number $2$.
Assume that the leaf with key $5$ gets deleted, 
so the node with key $1$ is now the left child 
of the node with key $5$, and node $1$'s prev pointer 
points to its former parent with key $3$.
Consider a \Find($0$, $1$) 
that has arrived at node $5$.  
It would go left from the node with key $5$ to the node with key $1$, 
but because the node with key $1$ has sequence number $2$, 
\Search\ would follow its $prev$ pointer
to the node with key $3$. 
Then it will go left to the node with key $1$ again
and this will repeat forever, thus creating an infinite loop. 
}

Finally, we discuss \Execute\ and \Help.
\Execute\ checks whether there are operations in progress
on the nodes that are to be flagged or marked and helps them if necessary
(lines~\ref{execute-for}--\ref{execute-false}). If this is not
the case, it creates a new Info object (line~\ref{create-info}), performs the first
flag \CAS\ to make the Info object visible to other processes (line~\ref{flagCAS1})
and calls \Help\ to perform the remaining actions (line~\ref{execute-help-self}).  
\Help($infp$) first performs the handshaking (line \ref{help-handshaking}--\ref{tryCAS}).
If $op$ does not abort (line~\ref{help-check-try1}), 
\Help\ attempts to flag and mark the remaining nodes recorded in the Info object pointed to by $infp$
(lines~\ref{help-check-try1}--\ref{help-while-end}). 
If it succeeds (line~\ref{help-check-continue}),
it executes a child \CAS\ to apply the required change on the appropriate tree pointer (line~\ref{help-CAS-child}).
If the child \CAS\ is successful, $op$ commits (line~\ref{commitWrite}),
otherwise it aborts (line~\ref{abortWrite}). 

\ignore{
Giati den bgainoun oi updates se palioteres ekdoseis tou dendrou wste na kanoun telika
update obsolete merh tou dendrou (mallon giati h Find and h Scan boithoun wste na teleiwnoun
oles autes h' na ginontai abort egkaira prin katalhksoun na doun kombous poy exoun megalytero
sequence number). 

++++++++++++
}

\section{Proof of Correctness}

\subsection{Proof Outline}

We first prove each call to a subroutine satisfies its preconditions.
This is proved together with some simple invariants, for instance, that \ReadChild($- , -, seq$)\
returns a pointer to a node whose sequence number is at most $seq$.
Next, we prove that $update$ fields of nodes
are updated in an orderly way and we study properties of the child \CAS\ steps.
A node $v$ is \emph{frozen for an Info object $I$} if 
$v.update$ points to $I$ and a call to \Frozen($v.update$)
would return \TRUE. A freeze \CAS\ (i.e., a flag or mark \CAS) {\em belongs to an Info object $I$}
if it occurs in an instance of \Help\ whose parameter is a pointer to $I$, 
or on line \lref{flagCAS1} with $I$ being the Info object created on line~\lref{create-info}.
We prove that only the first freeze \CAS\ that belongs to an Info object $I$ on each of the nodes in $I.nodes$
can be successful. 
Only the first child \CAS\ belonging to $I$ can succeed 
and this can only occur after all nodes in $I.nodes$ have been frozen.
If a successful child \CAS\ belongs to $I$, 
the $status$ field of $I$ never has the value \abort.
Specifically, this field is initially $\bot$ and changes to 
\try\ or \abort\ (depending on whether handshaking is performed successfully
on lines~\lref{help-handshaking}-\lref{tryCAS}).  If it changes to \try, 
then it may become \commit\ or \abort\ later (depending
on whether all nodes in $I.nodes$ are successfully frozen for $I$).
A node remains frozen for $I$ until $I.status$
changes to \commit\ or \abort. Once this occurs, the value of $I.status$ never changes again.
Only then can the $update$ field of the node
 become frozen for a different Info object. 
Values stored in $update$ fields of nodes
and in $child$ pointers are distinct (so no ABA problem may arise).

An {\em ichild} ({\em dchild}) \CAS\ is a child \CAS\ 
belonging to an Info object that was created 
by an \func{Insert} (\func{Delete}, respectively).
Note that executing a successful freeze \CAS\ (belonging to an Info object $I$
with sequence number $seq$) on a node $v$  
acts as a ``lock'' on $v$ set on behalf of the operation that created $I$. 
A successful child \CAS\ belonging to $I$
occurs only if the nodes that it will affect have been frozen.
Every such node has sequence number less than or equal to $seq$.
The ichild \CAS\ replaces a leaf $\ell$ with sequence number $i \leq seq$
with a subtree consisting of an internal node $v$ and two leaves (see Figure \ref{ins-example}).
All three nodes of this subtree have sequence number $seq$ and have never been in the tree before. 
Moreover, the $prev$ pointer of the internal node of this subtree points to $\ell$ 
(whereas those of the two leaves point to $\bot$). 
These changes imply that the execution of the ichild \CAS\ does not affect
any of the trees $T_i$ with $i < seq$. 
The part of the tree on which the ichild \CAS\ is performed
cannot change between the time all of the freeze \CAS\ steps (for $I$) were performed 
and the time the ichild \CAS\ is executed. So, the change that the 
ichild \CAS\ performs is visible in every $T_i$ with $i \geq seq$
just after this \CAS\ has been executed.
Similarly, a dchild \CAS\ does not cause any change to
any tree $T_i$ with $i < seq$. However, for each $i \geq seq$,
it replaces a node in $T_i$ with a copy of the sibling of the 
node to be deleted (which is a leaf), thus removing three nodes from the tree (see Figure \ref{ins-example}).

Characterizing the effects of child \CAS\ steps in this way allows us 
to prove that no node in $T_i$, $i \geq 0$, ever acquires a new ancestor after it is
first inserted in the tree. 
Using this, we also prove that if a node $v$ is in the search path 
for key $k$ in $T_i$ at some time, then it remains in the search
path for $k$ in $T_i$ at all later times.
We also prove that for every node $v$ an instance of \Search($k$, $seq$)\ traverses,
$v$ was in $T_{seq}$ (and on the search path for $k$ in it) at some time 
during the \func{Search}.
These facts allows us to prove that every $T_i$, $i \geq 0$, is a BST at all times.
Moreover, we prove that our validation scheme ensures that all successful update operations are applied
on the latest version of the tree. 

Fix an execution $\alpha$.
An update is {\em imminent} 
at some time during $\alpha$ if it has sucessfully executed
its first freeze \CAS\ before this time
and it later executes a successful child \CAS\ in~$\alpha$. 
We prove that at each  time,
no two imminent updates have the same key. 
For configuration $C$, let $Q(C)$
be the set of keys stored in leaves of $T_\infty$ at $C$
{\em plus} the set of keys of imminent \Insert\ operations at $C$
{\em minus} the set of keys of imminent \Delete\ operations at~$C$.
Let the {\em abstract set} $L(C)$ 
be the set that would result if all update operations with  
linearization points at or before $C$  
would be performed atomically 
in the order of their linearization points.
We prove the invariant that $Q(C)=L(C)$.
Once we know this, we can prove that each operation
returns the same result as it would if the operations were executed sequentially 
in the order defined by their linearization points, 
to complete the linearizability argument.

A \Scan\ with sequence number $i$ is wait-free because it traverses $T_i$, which
can only be modified by updates that begin before the \Scan's increment of the $Counter$ (due to handshaking).
To prove that the remaining operations are non-blocking, we show that an attempt of
an update that freezes its first node can only be blocked by an update that freezes a lower
node in the tree, so the update operating at a lowest node in the tree makes progress.

\subsection{Formal Proof}

We now provide the full proof of correctness.
Specifically, we prove that the implementation is linearizable and
satisfies progress properties.
The early parts of the proof are similar to proofs in previous 
work \cite{BER13,techrept,Sha13}, but are included here for completeness since the details differ.  
Most of the more novel aspects of the proof are in Sections
\ref{sec:tree properties} and \ref{sec:linearizability-argument}.

%

\later{when proof is complete, go through and double check that all lemmas (and all parts of lemmas)
are actually used later in the proof.}

\subsubsection{Basic Invariants}

We start by proving some simple invariants, and showing that there are no null-pointer exceptions in the code.

\begin{observation}
\label{unchanging}
The $key$, $prev$ and $seq$ fields of a Node never change. 
No field of an \Info\ \record, other than $state$, ever changes. 
The $Root$ pointer never changes.
\end{observation}

\begin{observation}\label{info-state}
If an Info object's state field is \commit\ or \abort\ in some configuration, it can never be $\bot$ or \try\ in a subsequent configuration.
\end{observation}
\begin{proof}
The state of an Info object can be changed only on lines \lref{abortCAS}, \lref{tryCAS}, \lref{commitWrite} and \lref{abortWrite}.  None of these can change the value from \commit\ or \abort\ to $\bot$ or \try.
\end{proof}

\begin{observation}
\label{seq-bound}
The value of $Counter$ is always non-negative, and
for every configuration $C$ and every node $v$ in configuration $C$, $v.seq \leq Counter$.
\end{observation}
\begin{proof}
The $Counter$ variable is initialized to 0 and never decreases.  All nodes in the initial configuration have $seq$ field 0.  Whenever a node is created by an \func{Insert} or \func{Delete}, its $seq$ field is assigned a value that the update operation read from $Counter$ earlier.
\end{proof}

\later{Technically, some of the proofs in the following should be enhanced to show that the node passed to various routines is of the right type (internal vs leaf).  For example, there was an error in the code (now fixed) where we were reading the children of a leaf if $sibling$ was a leaf during delete.}

\begin{invariant}
\label{simple-inv}
The following statements hold.
\begin{enumerate}
\item
\label{precond}
Each call to a routine satisfies its preconditions.
\item
\label{s-l-not-null}
Each \func{Search} that has executed line \lref{search-initialize} has local variables that satisfy the following:  $l \neq \NULL$ and  $l\rightarrow seq \leq seq$.
\item
\label{s-p-not-null}
Each \func{Search} that has executed line \lref{search-advance-p} has local variables that satisfy the following:  $p\neq\NULL$ and $p\rightarrow seq \leq seq$.
\item
\label{s-gp-not-null}
Each \func{Search} that has executed line \lref{search-initialize} has local variables that satisfy the following: if  $l\rightarrow key$ is finite then $gp\neq\NULL$ and $gp\rightarrow seq \leq seq$.
\item
\label{rc-l-not-null}
Each \func{ReadChild} that has executed line \lref{read-child} has local variables that satisfy the following:  $l\neq \NULL$ and there is a chain of $prev$ pointers from $l$ to a node whose $seq$ field is at most $seq$.
\item
\label{rc-post-seq}
Each \func{ReadChild} that terminates returns a pointer to a node whose sequence number is at most $seq$.
\item
\label{find-not-null}
Each \func{Find} that has executed line \lref{find-search} has non-\NULL\ values in its local variables $p$ and $l$.
\item
\label{insert-not-null}
Each \func{Insert} that has executed line \lref{insert-search} has local variables that satisfy the following:  $p\neq \NULL$ and $l\neq \NULL$ and $p\rightarrow seq \leq seq$.
\item
\label{delete-not-null}
Each \func{Delete} that has executed line \lref{delete-search} has local variables that satisfy the following:
$p\neq \NULL$ and $l\neq \NULL$ and $p\rightarrow seq\leq seq$.  Moreover, if $l\rightarrow key = k$, then $gp\neq\NULL$ and $gp\rightarrow seq\leq seq$.
\item
\label{child-defined}
For each Internal node $v$, $v$'s children pointers are non-\NULL.
Moreover, one can reach a node with sequence number at most $v.seq$ by tracing $prev$ pointers from either of $v$'s children.
\item
\label{info-inv}
For each Info object $I$ except $Dummy$, all elements of
$I.nodes$ are non-\NULL, $I.mark$ is a subset of $I.nodes$, $I.par$ is an element of $I.nodes$, $I.oldChild$ and $I.newChild$ are distinct and non-\NULL, $I.oldChild$ is an element of $I.mark$, and $I.newChild\rightarrow prev = I.oldChild$.
\item
\label{update-inv}
Each Update record has a non-\NULL\ $info$ field.
\item
\label{children-ordered}
For any Internal node $v$, any node $u$ reachable from $v.left$ by following a chain of $prev$ pointers has $u.key < v.key$
and any node $w$ reachable from $v.right$ by following a chain of $prev$ pointers has $w.key \geq v.key$.
\item
\label{root-newchild}
For any Info object $I$, if $I.par = Root$, then $I.newChild\rightarrow key$ is infinite.
\item
\label{root-left}
Any node $u$ that can be reached from $Root\rightarrow left$ by following a chain of $prev$ pointers has an infinite key.
\item
\label{readchild-result}
For any Internal node $v$, any terminating call to \func{ReadChild}$(v,\LEFT,seq)$ returns a node whose key is less than $v.key$, and any terminating call to \func{ReadChild}$(v,\RIGHT,seq)$ returns a node whose key is greater than or equal to $v.key$.  Any call to \func{ReadChild}$(Root,\LEFT,seq)$ returns a node whose key is infinite.
\end{enumerate}
\end{invariant}

\begin{proof}
We prove that all claims are satisfied in every finite execution by induction on the number of steps in the execution.

For the base case, consider an execution of 0 steps.  
Claims \ref{precond} to \ref{delete-not-null} are satisfied vacuously.
The initialization ensures that claims \ref{child-defined} to \ref{root-left} are true in the initial configuration.

Assume the claims hold for some finite execution $\alpha$.  We show that the claims hold for $\alpha \cdot s$, 
where $s$ is any step.
\begin{enumerate}
\item
If $s$ is a call to \func{Search} at line \lref{find-search}, \lref{insert-search} or \lref{delete-search}, the value of $seq$ was read from $Counter$ in a previous line.  The value of $Counter$ is always non-negative, so the precondition of the \func{Search} is satisfied.

If $s$ is a call to \func{ReadChild} on line \lref{search-advance-l}, the preconditions are satisfied by induction hypothesis \ref{s-p-not-null}.  
If $s$ is a call to \func{ReadChild} on line \lref{read-sibling}, the preconditions are satisfied by induction hypothesis \ref{delete-not-null}.
If $s$ is a call to \func{ReadChild} on line \lref{scanhelper-recursive1} to \lref{scanhelper-recursive4}, the preconditions are satisfied because \func{ScanHelper}'s preconditions were satisfied (by induction hypothesis \ref{precond}).

If $s$ is a call to \func{ValidateLink} on line \lref{validate-leaf-p} or \lref{validate-gp-p} of \ValidateLeaf, the preconditions follow from the preconditions of \ValidateLeaf, which are satisfied by induction hypothesis \ref{precond}.  (In the latter case, we know from the test on line \lref{validate-gp-p} that $p\neq \bot$.)
If $s$ is a call to \func{ValidateLink} on line \lref{delete-validate-p-sib}, the preconditions are satisfied
because the \Search\ on line \lref{delete-search} returned a node $p$ with sequence number at most $seq$
by induction hypothesis \ref{s-p-not-null}, and then \ReadChild\ on line \lref{read-sibling} returned
a node, by induction hypothesis \ref{rc-post-seq}.
If $s$ is a call to \func{ValidateLink} on line \lref{validate-sib-nephew1} or \lref{validate-sib-nephew2}, the 
preconditions are satisfied by induction hypothesis \ref{rc-post-seq} applied to the preceding call to \func{ReadChild} on line \lref{read-sibling}.

If $s$ is a call to \ValidateLeaf\ on line \lref{find-validate-leaf}, \lref{insert-validate-leaf} or \lref{delete-validate-leaf}, then the preconditions follow from induction hypotheses \ref{s-l-not-null}, \ref{s-p-not-null}, \ref{s-gp-not-null} and \func{readchild-result} applied to the preceding call to \func{Search} on line \lref{find-search}, \lref{insert-search} or \lref{delete-search}, respectively.

If $s$ is a call to \func{Execute} on line \lref{insert-execute} of \func{Insert}, preconditions (a)--(f) follow from induction hypothesis \ref{insert-not-null} and the fact that line \lref{create-internal} creates $newInternal$ after reading $l$ and sets  $newInternal\rightarrow prev$ to $l$.
It remains to prove precondition (g).  Suppose $p=Root$.  Since \func{ValidateLeaf} on line \lref{insert-validate-leaf} returned \func{True}, the call to \ValidateLink\ on line \lref{validate-leaf-p} also returned \TRUE.  So, $l$ was the result of the \func{ReadChild}$(Root, \lft, seq)$ on line \lref{val-read-child} of \func{ValidateLink}.  By induction hypothesis \ref{readchild-result}, $l$ has an  infinite key.  Thus, the new 
Internal node created on line \lref{create-internal} of the \func{Insert} has an infinite key, as required to satisfy precondition (g).

If $s$ is a call to \func{Execute} on line \lref{delete-execute} of \func{Delete}, 
preconditions (a)--(c) follow from induction hypothesis \ref{delete-not-null} and the fact that $l\rightarrow key = k$ (since the \func{Delete} did not terminate on line \lref{delete-false}), and induction hypothesis \ref{rc-post-seq} applied to the preceding call to \func{ReadChild} on line \lref{read-sibling}.
Precondition (d) follows from the additional fact that $newNode$ is created on line \lref{copy-sibling} after reading a pointer to $sibling$, which as already argued is non-\NULL.  Precondition (e) is obviously satisfied.
Precondition (f) follows from the fact that line \lref{copy-sibling} sets $newNode\rightarrow prev$ to be $p$.
It remains to prove precondition (g).  Suppose $gp=Root$.  
Since \func{ValidateLeaf} on line \lref{delete-validate-leaf} returned \func{True}, the call to \ValidateLink\ on line \lref{validate-gp-p} also returned \TRUE.  Then, $p$ was the result of the \func{ReadChild}$(Root, \LEFT, seq)$ on line \lref{val-read-child} of \func{ValidateLink}.  By induction hypothesis \ref{readchild-result}, $p$ has an  infinite key.
The \func{ReadChild}$(p, \RIGHT, seq)$ on line \lref{read-sibling} returns $sibling$, which also has an infinite key by induction hypothesis \ref{readchild-result}.  Thus, the node $newNode$ created at line \lref{copy-sibling} has an infinite key, as required to satisfy precondition (g).

If $s$ is a call to \func{Help} on line \lref{val-help}, \lref{execute-help-others} or \lref{scanhelper-help}, the argument is non-\NULL, by induction hypothesis \ref{update-inv}.
Moreover, the preceding call to \func{InProgress} returned true, so the Info object had state $\bot$ or \try.  By Observation \ref{info-state}, this Info object cannot be the Dummy object, which is initialized to have state \abort.
If $s$ is a call to \func{Help} on line \lref{execute-help-self}, the precondition is satisfied, since the argument $infp$ is created at line~\lref{create-info}.

If $s$ is a call to \func{CAS-Child} on line \lref{help-CAS-child}, the Info object $infp$ is not the Dummy, by the precondition to \func{Help}, which was satisfied when \func{Help} was called, by induction hypothesis \ref{precond}.
So, the preconditions of \func{CAS-Child} are satisfied by induction hypothesis \ref{info-inv}. 

If $s$ is a call to \func{ScanHelper} on line \lref{scan-return}, the precondition is satisfied since $Root\rightarrow seq = 0$ and the value of $Counter$ is always non-negative.
If $s$ is a call to \func{ScanHelper} on line \lref{scanhelper-recursive1} to \lref{scanhelper-recursive4}, the precondition is satisfied by induction hypothesis \ref{rc-post-seq}.

\item
By Observation \ref{unchanging}, the $seq$ field of a node does not change.  So it suffices to prove that any update to $l$ in the \func{Search} routine preserves the invariant.

Line \lref{search-initialize} sets $l$ to $Root$ which has $Root\rightarrow seq = 0$.
By induction hypothesis \ref{precond}, the \func{Search} has $seq \geq 0$, so claim \ref{s-l-not-null} is satisfied.

Line \lref{search-advance-l} sets $l$ to the result of a \func{ReadChild}, so claim \ref{s-l-not-null} is satisfied by induction hypothesis \ref{rc-post-seq}.

\item
It suffices to prove that any upate to $p$ in the \func{Search} routine preserves the invariant.  Whenever $p$ is updated at line \lref{search-advance-p}, it is set to the value stored in $l$, so claim \ref{s-p-not-null} follows from induction hypothesis \ref{s-l-not-null}.

\item
First, suppose $s$ is the first step of a \func{Search} that sets $l$ so that $l\rightarrow key$ is finite.
Then $s$ is not an execution of line \lref{search-initialize}, because $Root$ never changes and has key $\infty_2$, by Observation \ref{unchanging}.  Likewise, $s$ is not the assignment to $l$ that occurs in the first execution of line \lref{search-advance-l}, since the \func{ReadChild} on that line (which terminates before $s$) would have returned a node with an infinite key, by induction hypothesis \ref{readchild-result}.  Thus, $s$ occurs after the second execution of line \lref{search-advance-gp}, which happens after the first execution of line \lref{search-advance-p}.  By induction hypothesis \ref{s-p-not-null}, the second execution of line \lref{search-advance-p} assigns a non-null value to $gp$, and $gp\rightarrow seq \leq seq$.

It remains to consider any step $s$ that assigns a new value to $gp$ (at line \lref{search-advance-gp}) after
the first time $l$ is assigned a node with a finite value.  As argued in the previous paragraph, this execution of line \lref{search-advance-gp} will not occur in the first two iterations of the \func{Search}'s while loop.  So the claim follows
from induction hypothesis \ref{s-p-not-null}.

\item
By Observation \ref{unchanging}, $prev$ fields are never changed.  Thus,
it suffices to show that any step $s$ that updates $l$ inside the \func{ReadChild} routine maintains this invariant.

If $s$ is a step that sets $l$ to a child of $p$ at line \lref{read-child}, the claim follows from induction hypothesis \ref{child-defined} applied to the configuration just before $s$.

If $s$ is an execution of line \lref{read-prev}, the claim is clearly preserved.

\item
If $s$ is a step in which \func{ReadChild} terminates, the claim follows from induction hypothesis \ref{rc-l-not-null} applied to the configuration prior to $s$.

\item
It suffices to consider the step $s$ in which the \func{Search} called at line \lref{find-search} terminates.
That \func{Search} performed at least one iteration of its while loop (since $Root$ is an Internal node).  
So, by induction hypotheses \ref{s-l-not-null} and \ref{s-p-not-null}, it follows that the values that \func{Search} returns, which the \func{Find} stores in $p$ and $l$, are not \NULL.

\item
It suffices to consider the step $s$ in which the \func{Search} called at line \lref{insert-search} terminates.
That \func{Search} performed at least one iteration of its while loop (since $Root$ is an Internal node).  
So, by induction hypotheses \ref{s-l-not-null} and \ref{s-p-not-null}, it follows that the values that \func{Search} returns, which the \func{Insert} stores in $p$ and $l$, are not \NULL\ and have $seq$ fields that are at most $seq$.

\item
It suffices to consider the step $s$ in which the \func{Search} called at line \lref{delete-search} terminates.
That \func{Search} performed at least one iteration of its while loop (since $Root$ is an Internal node).  
So, by induction hypotheses \ref{s-l-not-null} and \ref{s-p-not-null}, it follows that the values that \func{Search} returns, which the \func{Delete} stores in $p$ and $l$, are not \NULL\ and have $seq$ fields that are at most $seq$.
If $l\rightarrow key = k$, it follows from induction hypothesis \ref{s-gp-not-null} that the value \func{Search} returns, which the \func{Delete} stores in $gp$, is not \NULL\ and that $gp\rightarrow seq\leq seq$.

\item
By Observation \ref{unchanging}, $prev$ pointers are never changed.  Thus, it suffices to show that every step $s$ that changes a child pointer preserves this invariant.
Consider a step $s$ that changes a child pointer by executing a successful child \CAS\ (at line \lref{childCAS1} or \lref{childCAS2}).
By the precondition of \func{CAS-Child}, the new child pointer will be non-\NULL\ and this new child's $prev$ pointer will point to the previous child.  Since one could reach a node with $seq$ field at most $seq$ by following $prev$ pointers from the old child (by induction hypothesis \ref{child-defined}), this will likewise be true if one follows $prev$ pointers from the new child.
\later{Youla says to check this part of proof for new internal node created by insert}

\item
By Observation \ref{unchanging}, the $nodes, mark, par, oldChild$ and $newChild$ fields of an Info object never change.  Thus it is sufficient to consider the case where the step $s$ is the creation of a new Info object at line \lref{create-info} of the \func{Execute} routine.  Claim \ref{info-inv} for the new Info object follows from the fact that the preconditions of \func{Execute} were satisfied when it was invoked before $s$.

\item
We consider all steps $s$ that construct a new Update record.
If $s$ is an execution of line \lref{flagCAS1}, the $info$ field of the new Update record is $infp$, which is defined on the previous line to be non-\NULL.
If $s$ is an execution of line \lref{markCAS} or \lref{flagCAS2} in the \func{Help} routine, the $info$ field of the new Update record is $infp$, which is non-\NULL, since induction hypothesis \ref{precond} ensures that the preconditions of the \func{Help} routine were satisfied when it was called.
If $s$ is an execution of line \lref{create-internal}, the $update$ field of the newly created node is set to a new Update record, $\langle \flag, Dummy\rangle$, which has a non-\NULL\ info field.

\item
If $s$ is a step that creates  a new Internal node $v$ (by executing line \lref{create-internal}), $v$'s left and right children
are initialized to satisfy the claim.

By observation \ref{unchanging}, $key$ and $prev$ fields of nodes are never changed, so it suffices to consider steps that change a child pointer.
If $s$ is a step that changes $v$'s child pointer (by executing line \lref{childCAS1} or \lref{childCAS2} in the \func{CAS-Child} routine) from $old$ to $new$,
it follows from the test on line \lref{which-child} that the new child $new$ has a key that satisfies the claim.
Moreover, by induction hypothesis \ref{precond}, the precondition of \func{CAS-Child} was satisfied when it was called, so  $new\rightarrow prev =old$.
By induction hypothesis \ref{children-ordered}, every node reachable from $old$ by following $prev$ pointers satisfied the claim.  So every node reachable from $new$ by following $prev$ pointers satisfies the claim too.

\item
By Observation \ref{unchanging}, an Info object's $par$ and $newChild$ fields do not change, and $prev$ and $key$ fields of nodes do not change.  Thus, it suffices to consider steps $s$ that create a new Info object (at line \lref{create-info} of the \func{Execute} routine).  The claim follows from the fact that the preconditions of \func{Execute} were satisfied when it was called, by induction hypothesis \ref{precond}.

\item
By observation \ref{unchanging}, $key$ and $prev$ fields of nodes are never changed, so it suffices to consider steps that change the left child pointer of $Root$.
Suppose $s$ is a step that changes $Root\rightarrow left$ (by executing line \lref{childCAS1} or \lref{childCAS2} in the \func{CAS-Child} routine) from $old$ to $new$.
That \func{CAS-Child} was called at line \lref{help-CAS-child} of \func{Help}.
By induction hypothesis \ref{root-newchild}, $new$ has an infinite key.
Moreover, by induction hypothesis \ref{precond}, the precondition of \func{CAS-Child} was satisfied when it was called, so  $new\rightarrow prev =old$.
By induction hypothesis \ref{children-ordered}, every node reachable from $old$ by following $prev$ pointers has an infinite key.  So every node reachable from $new$ by following $prev$ pointers has an infinite key.

\item
Suppose $s$ is the step in which a call to \func{ReadChild} returns.
By induction hypothesis \ref{children-ordered} and \ref{root-left}, 
when the \func{ReadChild} executed line \lref{read-child},
every node reachable from $l$ by following a chain of $prev$ pointers had the required property.  By Observation \ref{unchanging}, $prev$ pointers do not change.  So, the node returned by \func{ReadChild} has the required property.
\end{enumerate}

\end{proof}

\begin{invariant}
\label{node-info-seq}
For each Info object $I$ and each $i$, $I.nodes[i]\rightarrow seq \leq I.seq$.
\end{invariant}

\begin{proof}
By Observation \ref{unchanging}, the $nodes$ and $seq$ fields of Info objects, and the $seq$ fields of nodes do not change.  So it suffices to show that the claim is true whenever a new info object $I$ is created
(at line \lref{create-info} of \func{Execute}).
The \func{Execute} creates $I$
using the $nodes$ and $seq$ parameters of the call to \func{Execute}, which is called at line \lref{insert-execute} or \lref{delete-execute}.

If \func{Execute} is called at line \lref{insert-execute} of an \func{Insert}, the $nodes$ parameter contains nodes returned from a call to \func{Search}$(k,seq)$.  The sequence numbers of these two nodes are at most $seq$, by Invariant \ref{simple-inv}.\ref{s-p-not-null} and 
\ref{simple-inv}.\ref{s-l-not-null}, respectively.

If \func{Execute} is called at line \lref{delete-execute} of a \func{Delete}, the $nodes$ parameter contains nodes returned from a call to \func{Search}$(k,seq)$ on line \lref{delete-search} and a call to \func{ReadChild} on line \lref{read-sibling}.
The sequence numbers of these four nodes are at most $seq$, by Invariant \ref{simple-inv}.\ref{s-gp-not-null}, 
\ref{simple-inv}.\ref{s-p-not-null}, \ref{simple-inv}.\ref{s-l-not-null} and \ref{simple-inv}.\ref{rc-post-seq}, respectively.
\end{proof}

\subsubsection{How the $update$ Fields are Changed}

The next series of lemmas describes how $update$ fields of nodes are changed.  
This part of the proof is quite similar to other papers that have used similar techniques for flagging
or marking nodes, e.g., \cite{techrept,BER13}.  However, since we use a slightly different coordination
scheme from those papers, we include the lemmas here for the sake of completeness.

\begin{lemma}
\label{old-update-was-read}
For each Info object $I$ and all $i$, $I.oldUpdate[i]$ was read from the $update$ field of $I.nodes[i]$ prior to the creation of $I$.
\end{lemma}
\begin{proof}
Consider the creation of an Info object $I$ (at line \lref{create-info} of \func{Execute},
which is called either at line \lref{insert-execute} or \lref{delete-execute}).
So, it suffices to show that the claim is true for the arguments $nodes$ and $oldUpdate$ that are
passed as arguments in these calls to \func{Execute}.

If \func{Execute}$([p,l],[pupdate, l\rightarrow update], ...)$ was called at line \lref{insert-execute}, then $pupdate$ was read from $p\rightarrow update$ in the call to \func{ValidateLeaf} at line \lref{insert-validate-leaf}, and $l$'s update field is read at line \lref{insert-execute}.

If \func{Execute}$([gp,p,l,sibling],[gpupdate, pupdate, l\rightarrow update, supdate],...)$ 
was called at line \lref{delete-execute}, then $gpupdate$ and $pupdate$ were read from 
the $update$ fields of $gp$ and $p$ during the \func{ValidateLeaf} routine called at 
line \lref{delete-validate-leaf}.
The value of $supdate$ was read from $sibling\rightarrow update$ either during the call to 
\func{ValidateLink} at line \lref{validate-sib-nephew1} or at line \lref{read-sibling-update},
depending on whether $sibling$ is an Internal node or a Leaf.
Finally $l$'s update field is read at line \lref{delete-execute} itself.
\end{proof}

The following lemma shows that no ABA problem ever occurs on the $update$ field of a node.

\begin{lemma}
\label{update-no-ABA}
For each node $v$, the field $v.update$ is never set to a value that it has previously had.
\end{lemma}
\begin{proof}
The $v.update$ field can only be changed by the CAS steps at line \lref{flagCAS1}, \lref{markCAS} or \lref{flagCAS2}.
By Lemma \ref{old-update-was-read}, the CAS changes the $info$ subfield from a pointer to some Info object $I$ to a pointer to another Info object $I'$, where $I'$ was created after $I$.  The claim follows.
\end{proof}

We define some names for key steps for the algorithms that update the data structure.
The \CAS\ steps on lines \lref{flagCAS1} and \lref{flagCAS2} are called \emph{flag \CAS} steps, and the \CAS\ on line \lref{markCAS} is called a \emph{mark \CAS}.  
A \emph{freeze \CAS} step is either a flag \CAS\ or a mark \CAS.
An \emph{abort \CAS} occurs on line \lref{abortCAS} and a \emph{try \CAS} on line \lref{tryCAS}.  A \emph{child \CAS} occurs on line \lref{childCAS1} or \lref{childCAS2}.
Lines \lref{commitWrite} and \lref{abortWrite} are called \emph{commit writes} and \emph{abort writes}, respectively.

Any step performed inside a call to \func{Help}($infp$) is said to \emph{belong to} the Info object that $infp$ points to, including the steps performed inside the call to \func{\CAS-Child} on line \lref{help-CAS-child}.
The freeze \CAS\ on line \lref{flagCAS1} is also said to belong to the Info object created on the previous line.

\begin{lemma}
\label{only-first-freeze}
For each Info object $I$ and each $i$, only the first freeze \CAS\ on $I.nodes[i]$ that belongs to $I$ can succeed.
\end{lemma}
\begin{proof}
Let $u$ be the node that $I.nodes[i]$ points to.
All freeze \CAS\ steps on $u$ that belong to $I$ use the same old value $o$ for the \CAS, and $o$ is
read from $u.update$ prior to the creation of $I$.
If the first such freeze \CAS\ fails, then the value of $u.update$ has changed from $o$ to some other value before that first \CAS.
If the first freeze \CAS\ succeeds, then it changes $u.update$ to a value different from $o$ (since $o$ cannot contain a pointer to $I$ which was not created when $o$ was read, and the new value does contain a pointer to $o$).
Either way, the value of $u.update$ is different from $o$ after the first freeze \CAS, and it can never change
back to $o$ afterwards, by Lemma \ref{update-no-ABA}.
Thus, no subsequent freeze \CAS\ on $I.nodes[i]$ that belongs to $I$ can succeed.
\end{proof}

We next show that the $update$ field of a node can be changed only if the $state$
field of the Info object it points to is \commit\ or \abort.

\begin{lemma}
\label{not-still-trying}
Let $v$ be any node.  
If a step changes $v.update$, then $v.update.info\rightarrow state \in \{\commit, \abort\}$ in the configuration that precedes the step.
\end{lemma}
\begin{proof}
The only steps that can change $v.update$ are successful freeze CAS steps belonging to some Info object $I$ at line \lref{flagCAS1}, \lref{markCAS} or \lref{flagCAS2}.
Consider any such step $s$.  Since the freeze CAS succeeds, we have $v=I.nodes[i]$ for some $i$ and the value of $v.update$ prior to the step is $I.oldUpdate[i]$.  Let $I'$ be the Info object that $I.oldUpdate[i].info$ points to.
Prior to the creation of $I$ (at line \lref{create-info}), the call of \func{Frozen} on $I.oldUpdate[i]$ 
at line \lref{execute-check-frozen} returned \FALSE.
So, during the execution of line \lref{frozen}, $I'.state \in \{\commit, \abort\}$.
Once the state of $I'$ is either \commit\ or \abort, there is no instruction that can change it to $\bot$ or \try.
Thus, when $s$ occurs, $I'.state \in\{ \commit,\abort\}$, as required.
\end{proof}

\begin{lemma}
\label{abort-commit-exclusive}
If there is a child \CAS\ or commit write that belongs to an Info object $I$, then there is no abort write or successful abort \CAS\ that belongs to $I$.
\end{lemma}
\begin{proof}
Suppose there is a child \CAS\ or commit write that belongs to $I$.
Let $H$ be the instance of \func{Help} that performed this step.
At line \lref{help-check-try1} of $H$, $I.state$ was \try.
Thus, some try \CAS\ belonging to $I$ succeeded.
Let $try$ be this try \CAS.
Since there is no instruction that changes $I.state$ to $\bot$, this try \CAS\ must have been
the first among all abort \CAS\ and try \CAS\ steps belonging to $I$.  Moreover, no abort \CAS\ belonging to $I$ can ever succeed.

It remains to show that no abort write belongs to $I$.
To derive a contradiction, suppose there is such an abort write in some instance $H'$ of \func{Help}.
$I.state$ was \try\ when $H'$ executed line \lref{help-check-try2} prior to doing the abort write.
Since $try$ is the {\it first} among all try or abort \CAS\ steps belonging to $I$, $try$ is no
later than the execution of line \lref{abortCAS} or \lref{tryCAS} of $H'$.
Since no other step can change $I.state$ to \try, $I.state$ must have the value \try\ at all times between $try$ and the read by $H'$ at line \lref{help-check-try2}.  Thus, $H'$ reads $I.state$ to be \try\ at line \lref{help-check-try1} and sets the local variable $continue$ to \TRUE.
Since $H'$ executes the abort write at line \lref{abortWrite}, $H'$ must have
set continue to \FALSE\ at line \lref{help-check-frozen} after reading some value $I'$ different from $I$ in
$I.nodes[i]\rightarrow info$ for some~$i$.
Let $r$ be this read step.

Since $H$ performs a child \CAS\ or commit write belonging to $I$, $H$ must have read a pointer to $I$ in $I.nodes[i]\rightarrow update$ at line \lref{help-check-frozen}.  Thus, some freeze \CAS\, $fcas$ belonging to $I$ on $I.nodes[i]$ succeeded.
By Lemma \ref{only-first-freeze}, $fcas$ is the first freeze \CAS\ belonging to $I$ on $I.nodes[i]$.
So, $fcas$ is no later than the freeze \CAS\ of $H'$ on $I.nodes[i]$.
However, $I.nodes[i]\rightarrow update.info \neq I$ when $H'$ reads it on line \lref{help-check-frozen}.
So a successful freeze \CAS\ belonging to  $I'$ must have occurred between
$fcas$ and $r$.
By Lemma \ref{not-still-trying}, $I.state\in \{\commit,\abort\}$ when this successful freeze \CAS\ occurs.
This contradicts the fact that $I.state$ is still \try\ when $H'$ performs line \lref{help-check-try2}.
\end{proof}

\begin{corollary}
\label{irreversible}
Once an Info object's $state$ field becomes \abort\ or \commit, that
field can never change again.
\end{corollary}
\begin{proof}
No step can change the $state$ field to $\bot$.  It follows that no try \CAS\ can successfully change the $state$ field to \try, once it has become \commit\ or \abort.  Lemma \ref{abort-commit-exclusive} says that there cannot be two steps in the same execution that set the $state$ to \abort\ and \commit, respectively.
\end{proof}

We use the notation $\&X$ to refer to a pointer to object $X$.

\begin{lemma}
\label{done-trying}
At all times after a call $H$ to \func{Help}($\&I$) reaches line \lref{help-return}, the $state$ field of the Info object $I$ that $infp$ points to
is either \abort\ or \commit.
\end{lemma}
\begin{proof}
$I.state$ is initially $\bot$.
The first execution of line \lref{abortCAS} or \lref{tryCAS} belonging to $I$ changes the state to \abort\ or \try, and the state can never be changed back to $\bot$.
So, at all times after $H$ has executed line \lref{abortCAS} or \lref{tryCAS}, $I.state\neq \bot$.
If the condition at line \lref{help-check-continue} or \lref{help-check-try2} of $H$ evaluates to true, then $H$ writes \commit\ or \abort\ in $I.state$ at line \lref{commitWrite} or \lref{abortWrite}, respectively.  If both conditions evaluate to false, then $I.state$ is either \commit\ or \abort\ at line \lref{help-check-try2}.
In all three cases, $I.state$ has been either \commit\ or \abort\ at some time prior to $H$ reaching line \lref{help-return}.  The claim follows from Corollary \ref{irreversible}.
\end{proof}

%
%

\begin{lemma}
\label{freeze1-happened}
Let $I$ be an Info object other than the dummy Info object.  Let $C$ be any configuration.
If either
\begin{itemize}
\item
there is some node $v$, such that $v.update.info$ contains a pointer to $I$ in $C$, or
\item
some process is executing \func{Help}$(\&I)$ in $C$,
\end{itemize}
then there was a successful freeze \CAS\ at line \lref{flagCAS1} belonging to $I$
prior to $C$.
\end{lemma}
\begin{proof}
We prove this by induction on the length of the execution that leads to configuration $C$.
If $C$ is the initial configuration, the claim is vacuously satisfied.

Now consider any other configuration $C$ and assume the claim holds for all earlier configurations.
It suffices to show that any step $s$ that changes a node's $update$ field or invokes \func{Help} preserves the claim.

If $s$ is an invocation of \func{Help} at line \lref{execute-help-self} then it was clearly preceded by the freeze \CAS\ at line \lref{flagCAS1}.
If $s$ is an invocatino of \func{Help} at line \lref{val-help} or \lref{scanhelper-help}, then a pointer to $I$ was read
from a node's $update$ field at line \lref{val-read-pupdate} or \lref{scanhelper-read-info}, respectively,
so by the induction hypothesis, the claim holds.
If \func{Help} was called at line \lref{execute-help-others}, 
a pointer to $I$ appeared in a node's $update$ field in an earlier configuration by Lemma \ref{old-update-was-read}.
So, the claim again follows from the induction hypothesis.

If $s$ is an execution of line \lref{flagCAS1} itself that stores $I$ in some node's $update$ field, the claim is obvious.
If $s$ is an execution of line \lref{markCAS} or \lref{flagCAS2} of \func{Help}, then
the claim follows from the induction hypothesis (since a process was executing $\func{Help}(\&I)$ in the configuration preceding $s$).
\end{proof}

We next show that the freeze \CAS\ steps belonging to the same Info object occur in the right order.

\begin{lemma}
\label{freeze-in-order}
Let $I$ be an Info object.
For each $i\geq 2$, a freezing \CAS\ belonging to $I$ on $I.nodes[i]$ can occur only 
after a successful freezing \CAS\ belonging to $I$ on $I.nodes[i-1]$.
\end{lemma}
\begin{proof}
For $i=2$, since the freezing \CAS\ belonging to $I$ on $I.nodes[2]$ occurs inside \func{Help},
the claim follows from Lemma \ref{freeze1-happened}.

If $i>2$, then prior to the freezing \CAS\ on $I.nodes[i]$ at line \lref{markCAS} or \lref{flagCAS2},
$I.nodes[i-1]\rightarrow update.info$ contains a pointer to $I$ when line \lref{help-check-frozen} is executed
in the previous iteration of \func{Help}'s while loop.
Only a successful freezing \CAS\ on $I.nodes[i-1]$ belonging to $I$ could have put that value there.
\end{proof}

\begin{lemma}
Let $I$ be an Info object.
A successful freeze \CAS\ belonging to $I$ cannot occur when $I.state = \abort$.
\end{lemma}
\begin{proof}
There are no freeze \CAS\ steps of the dummy Info object, by the preconditions to \func{Help}.
Consider any other Info object $I$.
When a freeze \CAS\ at line \lref{flagCAS1} is performed, $I.state = \bot$.
Consider a successful freeze \CAS\ $fcas$ that belongs to $I$ inside some call $H$ to \func{Help}.
Then the test at line \lref{help-check-try1} of that call evaluated to true prior to $fcas$, so there is a successful try \CAS\ that belongs to $I$.  Thus, there is no successful abort \CAS\ that belongs to $I$.
It remains to show that no abort write belonging to $I$ occurred before $fcas$.

To derive a contradiction, suppose there was an abort write belonging to $I$ prior to $fcas$.
By Lemma \ref{abort-commit-exclusive}, there is no commit write belonging to $I$.
Consider the first abort write $w$ belonging to $I$.  Let $H'$ be the call to \func{Help} that performs $w$.
Prior to $w$, any execution of line \lref{help-check-try1} would find $I.state = \try$.
Thus, $H'$ set $continue$ to \FALSE\ at line \lref{help-check-frozen} when reading $I.nodes[i]\rightarrow update.info$ for some $i$.  Let $r$ be this read.
By Lemma \ref{freeze-in-order}, this step is preceded by freeze \CAS\ steps belonging to $I$ on each of
$I.nodes[1..i]$.  By Lemma \ref{only-first-freeze}, $fcas$ cannot be a freeze \CAS\ on any of these nodes,
so $fcas$ is a freeze \CAS\ on $I.nodes[j]$ for some $j>i$.

By Lemma \ref{freeze-in-order}, there is a successful freeze \CAS\ $fcas'$ on $I.nodes[i]$ belonging to $I$ before
$fcas$.  By Lemma \ref{only-first-freeze}, that \CAS\ precedes the read $r$ by $H'$ of $I.nodes[i]\rightarrow update.info$.  Since that read does not find a pointer to $I$ in that field, some other \CAS\ must have
changed it between $fcas'$ and $r$.  
This contradicts Lemma \ref{not-still-trying}, since $r$ precedes $w$, the first time $I.state$ gets set to \abort.
\end{proof}

\begin{definition}
We say that a node $v$ is \emph{frozen for an Info object $I$} if either
\begin{itemize}
\item
$v.update$ contains \flag\ and a pointer to $I$, and $I.state$ is either $\bot$ or \try, or 
\item
$v.update$ contains \mk\ and a pointer to $I$, and $I.state$ is not \abort.
\end{itemize} 
\end{definition}

\begin{lemma}
\label{stay-frozen}
\begin{enumerate}
\item
If there is a successful flag \CAS\ on node $v$ that belongs to Info object $I$, then $v$ is frozen for $I$ at all configurations that are after that \CAS\ and not after 
any abort \CAS, abort write or commit write belonging to $I$.
\item
If there is a successful mark \CAS\ on node $v$ that belongs to Info object $I$, then $v$ is frozen for $I$ at all configurations that are after that \CAS\ and not after 
any  abort write belonging to $I$.
\end{enumerate}
\end{lemma}

\begin{proof}
\begin{enumerate}
\item
It follows from Lemma \ref{not-still-trying} that $v.update$ cannot change after the successful flag \CAS, until
an abort \CAS, abort write or commit write belonging to $I$.
\item
If there is a successful mark \CAS\ $mcas$ belonging to $I$ (at line \ref{markCAS}), then the state of $I$
was \try\ at line \lref{help-check-try1}.  Thus, there is no successful abort \CAS\ belonging to $I$.
So, $v.update$ does not change until a commit write or an abort write belonging to $I$ occurs, by Lemma \ref{not-still-trying}.  We consider two cases.

If there is an abort write belonging to $I$, then there is no commit write belonging to $I$, so $v$ remains frozen
for $I$ in all configurations that are after $mcas$ but not after any  abort write belonging to $I$.

If there is no abort write belonging to $I$, then the state of $I$ is never set to \abort.
It remains to show that no freeze \CAS\ ever changes $v.update$ after $mcas$ changes it to $\langle \mk, \&I\rangle$.
Note that no info object $I'$ can have $I'.oldUpdate[i] = \langle \mk, \&I\rangle$.  If there were such
an $I'$, then before the creation of $I'$ at line \lref{create-info}, the call to \func{Frozen}($\langle \mk, \&I\rangle$) on line \lref{execute-check-frozen} would have had to return \FALSE, meaning that $I.state = \abort$, which is impossible.
So, no freeze \CAS\ belonging to any Info object $I'$ can change $v.update$ from $\langle \mk, \&I\rangle$ to some other value.  Thus, $v$ remains frozen for $I$ at all times after $mcas$.
\end{enumerate}
\end{proof}

\begin{corollary}
Let $v$ be a node and $I$ be an Info object.
If, in some configuration $C$, $v.update.type = \mk$ and $v.update.info$ points to $I$ and
$I.state = \commit$ then $v$ remains frozen for $I$ in all later configurations.
\end{corollary}
\begin{proof}
Prior to $C$ there must be a mark \CAS\ that sets $v.update$ to $\langle \mk, \&I\rangle$.
Since $I.state = \commit$, there is no abort write belonging to $I$, by Lemma \ref{abort-commit-exclusive}.
So the claim follows from Lemma \ref{stay-frozen}.
\end{proof}

\subsubsection{Behaviour of Child CAS steps}

Next, we prove a sequence of lemmas that describes how child pointers are changed.
In particular, we wish to show that our freezing scheme ensures that the appropriate nodes
are flagged or marked when a successful child \CAS\ updates the tree data structure.
Once again, these lemmas are similar to previous work \cite{techrept,BER13}, but are included for the sake of
completeness.

\begin{lemma}
\label{new-child-unique}
No two Info objects have the same value in the $newChild$ field.
\end{lemma}
\begin{proof}
Each Info object is created at line \lref{create-info} of the \func{Execute} routine, 
and no call to \func{Execute} creates more than one Info object.  Each call to 
\func{Execute} (at line \lref{insert-execute} or \lref{delete-execute}) passes a node 
that has just been newly created (at line \lref{create-internal} or \lref{copy-sibling}, 
respectively) as the argument that will become the $newChild$ field of the Info object.
\end{proof}

\begin{lemma}
\label{child-no-ABA}
The following are true for every Info object $I$  other than the dummy Info object.
\begin{enumerate}
\item
\label{child-CAS-new}
A successful child \CAS\ belonging to $I$ stores a value that has never been stored in that location before.
\item
\label{newChild-is-new}
If no child \CAS\ belonging to $I$ has occurred, then no node has a pointer to $I.newChild$ in its child or prev fields.
\end{enumerate}
\end{lemma}
\begin{proof}
We prove the lemma by induction on the length of the execution.
In an execution of 0 steps, the claim is vacuously satisfied, since there are no Info objects other than the dummy Info object.
Suppose the claim holds for some finite execution.  We show that it holds when the execution is extended by one step $s$.

If $s$ creates an Info object (at line \lref{create-info}) of the \func{Execute} routine,
the node $newChild$ was created at line \lref{create-internal} or \lref{copy-sibling}
prior to the call to \func{Execute} at line \lref{insert-execute} or \lref{delete-execute}.
Between the creation of the node and the creation of the Info object, a pointer to the node
is not written into shared memory.

If $s$ creates a new node, it is the execution of line \lref{create-leaf}, \lref{create-internal} or
\lref{copy-sibling}.  We must show that none of these nodes contain pointers to $I.newChild$ in their
child or $prev$ fields, for any $I$ whose first child \CAS\ has not yet occurred.
Line \lref{create-leaf} creates a leaf whose $prev$ field is $\bot$.
Line \lref{create-internal} sets one child pointer to $newSibling$, which does not appear in any shared-memory
location prior to line \lref{create-internal}.  The other child pointer and the $prev$ field are set to nodes
that were obtained from earlier calls to \func{ReadChild} and hence read from a $prev$ or child field
earlier.  By  induction hypotheses ref{newChild-is-new}, they cannot be $I.newChild$ for any Info object
$I$ whose first child \CAS\ has not occurred.
Similarly, when the node is created on line \lref{copy-sibling}, its $prev$ and child fields are
set to values that were read from $prev$ or child fields of other nodes, so the same argument applies.

If $s$ is the first child \CAS\ belonging to $I$, claim \ref{child-CAS-new} follows from induction hypothesis \ref{newChild-is-new}.

If $s$ is not the first child \CAS\ belonging to $I$, we prove that it is not successful.
To derive a contradiction, suppose some earlier child \CAS\ $s'$ belonging to $I$ also succeeded.
Both $s$ and $s'$ perform $\CAS(location, old, new)$ steps with identical arguments.
Thus $location$ stores the value $old$ in the configurations just before $s'$ and $s$ (since both \CAS\ steps
succeed).
By Lemma \ref{simple-inv}.\ref{info-inv}, $old \neq new$.  So, between $s'$ and $s$, there must be some child \CAS\ that changes
$location$ from $new$ back to $old$.  This violates part \ref{child-CAS-new} of the inductive hypothesis.

If $s$ is a child \CAS\ belonging to some other Info object $I'\neq I$, then it does not write a pointer
to $I.newChild$ into any node, by Lemma \ref{new-child-unique}.
\end{proof}

\begin{corollary}
\label{only-first-child-CAS}
Only the first child \CAS\ belonging to an Info object can succeed.
\end{corollary}
\begin{proof}
Since all child \CAS\ steps belonging to the same Info object try to write the same value into the same location,
only the first can succeed, by Lemma \ref{child-no-ABA}.\ref{child-CAS-new}.
\end{proof}

\begin{lemma}
\label{frozen-at-child-CAS}
The first child \CAS\ belonging to an Info object $I$ occurs while all nodes in $I.nodes$ are frozen for $I$, including the node $I.par$ to which the child \CAS\ is applied.
\end{lemma}
\begin{proof}
Since there is a child \CAS\ belonging to $I$, there is no abort write or successful abort \CAS\ belonging to $I$, by Lemma \ref{abort-commit-exclusive}.
Prior to the call to \func{CAS-Child} on line \lref{help-CAS-child} that performed the successful child \CAS,
the local variable $continue$ was true at line \lref{help-check-continue}.
This means that a freeze \CAS\ belonging to $I$ succeeded on each entry of $I.nodes[i]$, including $I.par$, by Lemma \ref{simple-inv}.\ref{info-inv}.
By Lemma \ref{stay-frozen}, these nodes remain frozen for $I$ in all configurations that are after that freeze
\CAS\ and not after a commit write belonging to $I$.
The first child \CAS\ that belongs to $I$ is before the first commit write belonging 
to $I$.
So, the nodes in $I.nodes$ (including $I.par$) are frozen for $I$ when this  child \CAS\ occurs.
\end{proof}

The following lemma shows that marking a node is permanent, if the attempt of the update that marks the node succeeds.

\begin{lemma}
\label{marking-permanent}
If there is a child \CAS\ belonging to an Info object $I$, then for all $i$,
$I.mark[i]\rightarrow update = \langle \mk, \&I\rangle$ in all configurations after the first such child \CAS.
\end{lemma}
\begin{proof}
By Lemma \ref{frozen-at-child-CAS}, the claim is true in the configuration immediately after the first child
\CAS\ belonging to $I$.
To derive a contradiction, suppose the $update$ field of $I.mark[i]$ is later changed.
Consider the first such change.
This change is made by a successful freezing \CAS\ belonging to some Info object $I'$.
Before $I'$ is created at line \lref{create-info},
\func{Frozen}$(\langle \mk, \&I\rangle)$ returns \FALSE\ at line \lref{execute-check-frozen},
so $I.state = \abort$.  This contradicts Lemma \ref{abort-commit-exclusive}.
\end{proof}

The next lemma shows that if at some time the $update$ field of a node $v$ has the value $I.oldupdate[i]$ for some Info object $I$
and at some later time $v$ is still frozen for $I$ then a child pointer of $v$ can change between these times
only by a successful child \CAS\ that belongs to $I$.  (Thus, the freezing works as a `lock' on the child
pointers of the node.)

\begin{lemma}
\label{no-child-changes}
Let $I$ be an \Info\ object and let $v$ be the node that $I.nodes[i]$ points to, for some $i$.
If $v.update = I.oldUpdate[i]$ in some configuration $C$ and $I.info\rightarrow state \in \{\commit,\abort\}$ in $C$, 
and $v$ is frozen for $I$ in a later configuration $C'$, then the only step between $C$ and $C'$ that might change a
child field of $v$  is a successful child \CAS\ belonging to $I$.
\end{lemma}

\begin{proof}
Since $v.update = I.oldUpdate[i]$ at configuration $C$, and $v.update = \langle *, I\rangle$ at configuration $C'$,
there is a successful freeze \CAS\ $fcas$ that belongs to $I$ on $v$ between $C$ and $C'$.
This freeze \CAS\ uses $I.oldUpdate[i]$ as the expected value of $v.update$.
So, by Lemma \ref{update-no-ABA}, $v.update = I.oldUpdate[i]$ at all configurations between $C$ and $fcas$,
and $v.update = \langle *, I\rangle$ at all times between $fcas$ and $C'$.  Let $I'$ be the Info
object that  $I.oldUpdate[i].info$ points to.

By Corollary \ref{only-first-child-CAS} and Lemma \ref{frozen-at-child-CAS}, 
any successful child \CAS\ on $v$ between $C$ and $C'$ must belong to either $I'$ or $I$.
To derive a contradiction, suppose there is such a successful child \CAS\ that belongs to $I'$.
Then by Lemma \ref{abort-commit-exclusive}, there is no abort \CAS\ or abort write that belongs to $I'$.
By Lemma \ref{only-first-child-CAS}, this successful child \CAS\ is the first child \CAS\ of $I'$,
which is before the first commit write belonging to $I'$.
Thus, $I'.state \notin \{\commit,\abort\}$ in $C$ because $C$ is before the successful child \CAS,
contradicting the hypothesis of the lemma.
\end{proof}

\begin{lemma}
\label{child-CAS-succeeds}
For any Info object $I$, the first child \CAS\ that belongs to $I$ succeeds.
\end{lemma}
\begin{proof}
Let $v$ be the node that $I.nodes[1]$ points to and let $u$ be the node that $I.oldChild$ points to.
The Info object $I$ is created at line \lref{create-info} of the \func{Execute} routine.
Before \func{Execute} is called at line \lref{insert-execute} or \lref{delete-execute},
there is a call to \func{ValidateLeaf} on line \lref{insert-validate-leaf} or \lref{delete-validate-leaf}, respectively.
\func{ValidateLeaf} calls \func{ValidateLink}, which returns \TRUE.
This \func{ValidateLink}
reads a value from  $v.update$ that is ultimately
stored in $I.oldUpdate[1]$ and then checks on line \lref{val-check-frozen} that
 $v.update.state \notin \{\bot,\try\}$ when $v.update$ was read on line \lref{val-read-pupdate}.
Let $C$ be the configuration after this read.
After $C$, on line \lref{val-read-child}, the value $u$ is read from a child field of $v$.

Let $C'$ be the configuration just before the first child \CAS\ belonging to $I$.
By Lemma \ref{frozen-at-child-CAS}, $v$ is frozen for $I$ in $C'$.
So, by Lemma \ref{no-child-changes}, there is no change to $v$'s child fields
between $C$ and  $C'$.
Moreover, $u$ is read from a child field of $v$ during this period, and the first child \CAS\ of $I$ uses
$u$ as the old value, so it will succeed.
\later{Technically, there should be some argument that the read is of the {\it same} child field
as the child \CAS}
\end{proof}

\subsubsection{Tree Properties}
\label{sec:tree properties}

In this section, we use the lemmas from the previous sections to begin proving higher-level
claims about our particular data structure, culminating in Lemma \ref{search-linearizable},
which proves that \func{Searches} end up at the correct leaf, and Lemma \ref{bst-inv}, which 
proves that all versions of the tree are BSTs.

Our data structure is persistent, so it is possible to reconstruct previous versions of it.
Consider a configuration $C$.
For any Internal node $v$ whose sequence number is at most $\ell$, we define the {\it version-$\ell$ left (or right) child of $v$} to be the node that is reached by following the
left (or right) child pointer of $v$ and then following its $prev$ pointers until reaching the first node whose $seq$ field
is less than or equal to $\ell$.  
(We shall show that such a node exists.) 
We define $D_{\ell}(C)$ as follows.
The nodes of $D_{\ell}(C)$ is the set of all existing nodes in $C$ and the edges go from nodes to their version-$\ell$ children;
$T_\ell(C)$ is the subgraph of $D_\ell(C)$ containing those nodes that are reachable from the $Root$ in $D_\ell(C)$.
We use the notation $T_\infty(C)$ to represent the graph of nodes reachable from the $Root$ by following the current child pointers.
We shall show that 
$T_\ell(C)$ is a binary search tree rooted at $Root$.

\begin{definition}
We say a node is \emph{inactive} when it is first created.
If the node is created at line \lref{create-internal} or \lref{copy-sibling}, it becomes active when
a child \CAS\ writes a pointer to it for the first time, and it remains active forever afterwards.
If the node is created at line \lref{create-new-leaf} or \lref{create-sibling-leaf}, then it becomes active
when a child \CAS\ writes a pointer to its parent for the first time, and it remains active forever afterwards.
The nodes that are initially in the tree are always active. 
\end{definition}

\begin{definition}
An ichild \CAS\ is a child \CAS\ belonging to an Info object that was created by an \func{Insert} and a dchild \CAS\ is a child
\CAS\ belonging to an Info object that was created by a \func{Delete}.
\end{definition}

\begin{lemma}
\label{active-inv}
\begin{enumerate}
\item
\label{pointer-to-active}
If a node is inactive, then there is no pointer to it in the $prev$ field of any node or in a child field of an active node.
\item
\label{argument-active}
The first argument of each call to \func{ReadChild} and \func{ScanHelper} is an active node.
\item
\label{return-active}
No call to \func{ReadChild} or \func{Search} returns an inactive node. 
\item
\label{info-active}
For each Info object $I$, $I.nodes$ contains only active nodes.
\end{enumerate}
\end{lemma}
\begin{proof}
We prove the claim by induction on the length of the execution.
The claim is vacuously satisfied for an execution of length 0.
Assume the claim holds for some execution.  We prove that it holds when the execution is extended by one step $s$.

\begin{enumerate}
\item
When the $prev$ field of a node is set at line \lref{create-internal}, it points to a node returned by the \func{Search} on line \lref{insert-search}, so it is active by inductive hypothesis \ref{return-active}.
When the $prev$ field of a node is set at line \lref{copy-sibling}, it points to a node returned by the
\func{ReadChild} on the previous line, which is active by inductive hypothesis \ref{return-active}.

If $s$ is a successful child \CAS\ that changes a child pointer to point to a node $v$, $v$ is active after the child \CAS, by definition.  If $v$ was created at line \lref{create-internal}, its children become active at the same time as $v$.  If $v$ was created at line \lref{copy-sibling}, any children it has were copied from the children fields of an active node by induction hypothesis \ref{pointer-to-active}, so they were already active when $v$ was created.

\item
If $s$ is a call to \func{ReadChild} on line \lref{search-advance-l}, the first argument is either the root node,
which is active, or the result of a previous call to \func{ReadChild}, which is active by inductive hypothesis \ref{return-active}.
If $s$ is a call to \func{ReadChild} on line \lref{read-sibling}, the first argument was returned by \func{Search} on line \lref{delete-search}, so it is active by inductive hypothesis \ref{return-active}.
If $s$ is a call to \func{ReadChild} on line \lref{scanhelper-recursive1} to \lref{scanhelper-recursive4},
then the first argument is the first argument of the call to \ScanHelper, so it is active by inductive hypothesis \ref{argument-active}.

If $s$ is a call to \func{ScanHelper} on line \lref{scan-return}, the first argument is the root node, which is active.
If $s$ is a call to \func{ScanHelper} on line \lref{scanhelper-recursive1} to \lref{scanhelper-recursive4},
the first argument was returned by a call to \func{ReadChild}, which was active by inductive hypothesis \ref{return-active}.

\item
Suppose $s$ is the return statement of a \func{ReadChild}.  When that function was called, the first
argument was an active node, by inductive hypothesis \ref{argument-active}.  The node returned by \func{ReadChild}
is reached from that node by following child and prev pointers, so it follows from inductive hypothesis \ref{pointer-to-active}
that the resulting node is active too.

Suppose $s$ is the return statement of a \func{Search}.  Each node returned is either the root, which is active, or obtained as the result of a \func{ReadChild} at line \lref{search-advance-l} during the \func{Search}, which is active by inductive hypothesis \ref{return-active}.

\item
Suppose $s$ is a step that creates an Info object $I$ at line \lref{create-info} of \func{Execute}.
If \func{Execute} was called at line \lref{insert-execute}, then the elements of $I.nodes$ were returned
by the \func{Search} on line \lref{insert-search}, so they are active by inductive hypothesis \ref{return-active}.
If \func{Execute} was called at line \lref{delete-execute}, then the elements of $I.nodes$ were returned
by the \func{Search} on line \lref{insert-search} or the \func{ReachChild} on line \lref{read-sibling},
so they are active by inductive hypothesis \ref{return-active}.
\end{enumerate}
\end{proof}

The following Lemma shows that the effect of a child \CAS\ step is as shown in Figure \ref{ins-example}.

\begin{lemma}
\label{child-cas-lemma}
Consider a successful child \CAS\ step $s$ that belongs to some Info object $I$.
Let $C$ and $C'$ be the configurations before and after $s$.
Then,
\begin{enumerate}
\item
In $C$, $I.oldChild \rightarrow update = \langle \mk, \&I\rangle$.
\item
In $C$, $I.newChild$ is inactive.
\item
\label{ichild-direct-effect}
If $s$ is an ichild \CAS\ created by an \func{Insert}($k$) then $I.newChild$ is an internal node and its two children in $C'$ are both leaves, one of which has the same key as $I.oldChild$ and the other has the key $k$.
\item
\label{dchild-direct-effect}
If $s$ is a dchild \CAS\ created by a \func{Delete}($k$) operation then $I.oldChild$ is an Internal node and in configuration $C$:
\begin{itemize}
\item one of its children is $I.nodes[3]$, which is a leaf containing the key $k$, and 
\item the other child is $I.nodes[4]$, which has the same key and children as $I.newChild$, and 
\item both of the children of $I.oldChild$ have $\langle\mk, \&I\rangle$ in their $update$ fields.  
\end{itemize}
\end{enumerate}
\end{lemma}

\begin{proof}
By Corollary \ref{only-first-child-CAS}, $s$ is the first child \CAS\ belonging to $I$.
\begin{enumerate}
\item
By Lemma \ref{simple-inv}.\ref{info-inv}, $I.oldChild$ is in $I.mark$, which is a subset of $I.nodes$.
So, by Lemma \ref{frozen-at-child-CAS}, $I.oldChild$ is frozen for $I$ at $C$ and it must have been a mark
\CAS\ that froze the node.

\item 
Note that $I.newChild$ was created at line \lref{create-internal} if $s$ is an ichild \CAS, 
or at line \lref{copy-sibling} if $s$ is a dchild \CAS.
By Lemma \ref{new-child-unique}, $s$ is the first child \CAS\ that writes a pointer to $I.newChild$, so 
this node becomes active for the first time in $C'$.

\item
$I.newChild$ was created at line \lref{create-internal}, with its children satisfying the claim,
and the children pointers cannot be changed before $I.newChild$ becomes active at $C'$, 
by Lemma \ref{active-inv}.\ref{info-active}.

\item
Since $s$ is a dchild \CAS, $I$ was created by an \func{Execute} routine called at line \lref{delete-execute}
of a \func{Delete}($k$) operation.
$I.oldChild$ is copied from the local variable $p$ of that \func{Delete}.
$I.oldUpdate[2]$ was read from the $update$ field of $I.oldChild$ inside the call at line \lref{delete-validate-leaf}.
Since that call to \func{ValidateLeaf} returned $\langle \TRUE, I.oldUpdate[2]\rangle$, $I.oldChild.info$ was found to be an Info object that was in state \abort\ or \commit.
Subsequently, the two child fields of $I.oldChild$ were read (inside the same call to \ValidateLeaf\ and at line \lref{delete-validate-p-sib}) and were seen to be equal to $l$ and $sibling$.  
By Lemma \ref{no-child-changes}, these are still the children of $I.oldChild$ in $C$ since $I.oldChild=I.nodes[2]$ is frozen for $I$ in $C$, by Lemma \ref{frozen-at-child-CAS}.
By the exit condition at line \lref{search-while} of the \func{Search} called at line \lref{delete-search}, $l$ is a leaf.
Furthermore, $l\rightarrow key = k$,
since the test at line \lref{delete-false} evaluated to \FALSE.

The key and children of $I.newChild$ are copied from $sibling$.  
If $sibling$ is a leaf, then $I.newChild$ is also a leaf, so there is nothing further to prove.
If $sibling$ is an internal node, it remains to prove that the children of $sibling$ do
not change between the time they are copied at line \lref{copy-sibling} and $C$.
This is because the call to \func{ValidateLink} (at line \lref{validate-sib-nephew1})
read $I.oldUpdate[4]$ from $sibling\rightarrow update$ and then sees that the Info object that field
points to is in state \commit\ or \abort.  Subsequently the children of $sibling$ are seen to
be the two children of $I.newChild$ inside the calls to \func{ValidateLink} at line \lref{validate-sib-nephew1}
and \lref{validate-sib-nephew2}.
By Lemma \ref{no-child-changes}, these are still the children of $sibling$ in $C$ since $sibling = I.nodes[4]$ is frozen for $I$ in $C$, by Lemma \ref{frozen-at-child-CAS}.

Both $l$ and $sibling$ are included in $I.nodes$.  
By Lemma \ref{frozen-at-child-CAS} they are both frozen for $I$ at configuration $C$.
Since they are also in $I.mark$, they were frozen for $I$ by a mark \CAS, so their update fields are $\langle \mk, \&I\rangle$.
\end{enumerate}
\end{proof}

By Observation \ref{unchanging}, no step changes a $prev$ pointer of an existing node.
The only step that changes a child field of a node is a successful child \CAS.
Thus, the following lemma provides a complete description of how $T_i$ can be changed by any step.
It also characterizes which nodes are in different tree versions $T_i$:  roughly speaking, 
if a node is flagged, then it is still in all versions of the tree, 
but if it is marked for removal, it will be in all versions of the tree if the corresponding child \CAS\ has not yet occurred, but it will only be in old versions after the child \CAS\ has removed it.

\later{The following proof is ugly, but it works.  Reorganize it later to make it prettier.  In particular the proof of 2 can probably be done so that (a),(b),(c) are all proved at once by considering all steps that could potentially make one of those three statements false. }


\begin{lemma}
\label{child-CAS-super-lemma}
The following statements hold.
\begin{enumerate}
\item
\label{child-CAS-effect}
For each successful child \CAS\ that belongs to some Info object $I$ and takes the system from configuration $C$ to $C'$, the following statements are true.
\begin{enumerate}
\item
\label{child-CAS-effect-low}
For all $i<I.seq$, $T_i(C) = T_i(C')$.
\item
\label{ichild-CAS-effect-high}
If $I$ was created by an \func{Insert}($k$), then
for all $i\geq I.seq$, $T_i(C')$ is obtained from $T_i(C)$ by replacing the leaf 
$I.oldChild$ by $I.newChild$, which is an internal node whose children are two leaves with keys $I.oldChild\rightarrow key$ and $k$.  (If $I.oldChild$ is not in $T_i(C)$, then this replacement has no effect on $T_i$.)
\item
\label{dchild-CAS-effect-high}
If $I$ was created by a \func{Delete}($k$), then
for all $i\geq I.seq$, $T_i(C')$ is obtained from $T_i(C)$ by replacing the internal node $I.oldChild$
and its two children (which are a leaf containing $k$ and a node $sibling$) by a copy 
of $I.newChild$, whose key is $sibling.key$ and whose children are the same as $sibling$'s children.
(If $I.oldChild$ is not in $T_i(C)$, then this replacement has no effect on $T_i$.)
\end{enumerate}
\item
\label{is-node-in-tree}
For every configuration $C'$, and for each node $v$ that is active in $C'$, and for all $i \geq v.seq$, the following statements are true.
\begin{enumerate}
\item \label{is-node-in-tree-flag}
If $v.update.type = \flag$ in $C'$ then $v$ is in $T_i(C')$.
\item \label{is-node-in-tree-mark1}
If $v.update = \langle \mk, \&I \rangle$ in $C'$ and no child \CAS\ that belongs to $I$ has occurred before $C'$, then $v$ is in $T_i(C')$.
\item \label{is-node-in-tree-mark2}
If $v.update = \langle \mk, \&I \rangle$ in $C'$ and $i<I.seq$, then $v$ is in $T_i(C')$.
\end{enumerate}
\end{enumerate}
\end{lemma}
\begin{proof}
We prove the claim by induction on the length of the execution.  First consider an execution of 0 steps.
Claim \ref{child-CAS-effect} is satisfied vacuously.  In the initial configuration $C_0$, all nodes are active, flagged with the dummy Info object, have sequence number 0, and are in $T_i(C_0)$ for all $i$, so claim \ref{is-node-in-tree} is true.  

Now, suppose the claim holds throughout some finite execution.  We prove the claim holds for any extension of that execution by a single step $s$.

\begin{enumerate}
\item 
Claim \ref{child-CAS-effect} for all successful child \CAS\ steps prior to $s$ follows from induction hypothesis \ref{child-CAS-effect}.
So it suffices to prove claim \ref{child-CAS-effect} holds for $s$ if $s$ is a successful child \CAS\ belonging to some Info object~$I$.
\begin{enumerate}
\item
When $I$ is created, $I.newChild$ is given the sequence number $I.seq$.
Thus, when $s$ swings a child pointer from $I.oldChild$ to $I.newChild$
it does not affect $T_i$ for $i<I.seq$, since $I.newChild \rightarrow prev = I.oldChild$, by Lemma \ref{simple-inv}.\ref{info-inv}.

\item
Suppose $I$ was created by an \func{Insert}($k$) operation.
Consider any $i\geq I.seq$.
The step $s$ changes a child pointer of some node $p$ from $I.oldChild$ to $I.newChild$.
By Lemma \ref{active-inv}.\ref{info-active}, $p$ is active in $C$, so we can apply induction hypothesis \ref{is-node-in-tree} to it.
By Lemma \ref{frozen-at-child-CAS}, $p$ is frozen for $I$ in $C$.
Since $p$ is not in $I.mark$, $p.update.type = \flag$.
Moreover, $i\geq I.seq \geq p.seq$ by Invariant \ref{node-info-seq}.
So, by induction hypothesis \ref{is-node-in-tree-flag}, $p$ is in $T_i(C)$.
Claim \ref{ichild-CAS-effect-high} follows from Lemma \ref{child-cas-lemma}.\ref{ichild-direct-effect}, since
$I.newChild\rightarrow seq = I.seq \leq i$.

\item
Suppose $I$ was created by a \func{Delete}($k$) operation.
Consider any $i\geq I.seq$.
The step $s$ changes a child pointer of some node $gp$ from $I.oldChild$ to $I.newChild$.
By Lemma \ref{active-inv}.\ref{info-active},  $gp=I.nodes[1]$ is active in $C$, so we can apply induction hypothesis \ref{is-node-in-tree} to it.
By Lemma \ref{frozen-at-child-CAS}, $gp$ is frozen for $I$ in $C$.
Since $gp$ is not in $I.mark$, $gp.update.type = \flag$.
Moreover, $i\geq I.seq \geq gp.seq$ by Invariant \ref{node-info-seq}.
So, by induction hypothesis \ref{is-node-in-tree-flag}, $gp$ is in $T_i(C)$.
Claim \ref{dchild-CAS-effect-high} follows from Lemma \ref{child-cas-lemma}.\ref{dchild-direct-effect}, since
$I.newChild\rightarrow seq = I.seq \leq i$.
\end{enumerate}
\item
Induction hypothesis \ref{is-node-in-tree} establishes the claim for all configurations prior to the final step $s$,
so it suffices to prove the claim for the configuration $C'$ after $s$.
Let $C$ be the configuration before $s$.
Let $v$ be any node that is active in $C'$ and let $i\geq v.seq$.

\begin{enumerate}
\item
Suppose $v.update.type=\flag$ in $C'$.
We consider four cases.
\begin{itemize}
\item
Suppose $s$ is the child \CAS\ that makes $v$ active. Then, by Lemma \ref{frozen-at-child-CAS}
the node $p$ whose child pointer is modified by $s$ is flagged in $C$.
Let $I$ be the Info object that $s$ belongs to.
By induction hypothesis \ref{is-node-in-tree-flag}, $p$ is in $T_i(C)$ since $i\geq v.seq = I.seq \geq p.seq$ by Lemma~\ref{node-info-seq}. 
So, the node $v$ is in $T_i(C')$ since there is now a path of child pointers from $p$ to $v$ of nodes whose sequence numbers are $v.seq$.

\item
Suppose $v$ is active in $C$ and $s$ is a successful flag \CAS\ on $v$.
Let $I$ be the Info object that $s$ belongs to.  
Let $up$ be the value stored in $v.update$ in $C$.

If $up.type = \flag$, then by induction hypothesis \ref{is-node-in-tree-flag}, $v$ was in $T_i(C)$, so it is in $T_i(C')$.

Now suppose $up = \langle\mk ,\&I'\rangle$ for some Info object $I'$.
Since $s$ belongs to $I$, $v=I.nodes[j]$ for some $j$
and $up = I.oldUpdate[j]$.
Prior to the creation of $I$ at line \lref{create-info}, the call to \func{Frozen}($up$)
at line \lref{execute-check-frozen} returned \FALSE.
So, the test at line \lref{frozen} found $I'.state = \abort$.
By Lemma \ref{abort-commit-exclusive}, there is no child \CAS\ belonging to $I'$.
So by induction hypothesis \ref{is-node-in-tree-mark1}, $v$ is in $T_i(C)$, so it is also in $T_i(C')$.

\item
Suppose $v$ is active in $C$ and $s$ is a successful child \CAS.
If $i<I.seq$, then $T_i(C) = T_i(C')$ (by claim \ref{child-CAS-effect-low} proved above), so claim \ref{is-node-in-tree-flag} follows from induction hypothesis \ref{is-node-in-tree-flag}.

Now suppose $i\geq I.seq$.
If $s$ is an ichild \CAS, then by claim \ref{ichild-CAS-effect-high}, proved above, the only node that $s$ removes from $T_i$ is $I.oldChild$, which is marked for $I$ in $C$ and is therefore not $v$ (since $v$ is flagged in $C$).  
If $s$ is a dchild \CAS, then by claim \ref{dchild-CAS-effect-high}, proved above, the only nodes that $s$ removes from $T_i$ are
$I.oldChild$ and its children.  By Lemma \ref{child-cas-lemma}.\ref{dchild-direct-effect}, these nodes are
the three nodes in $I.mark$.  So by lemma \ref{frozen-at-child-CAS}, they are marked in $C'$ and are therefore not equal to $v$.
In either case, claim \ref{is-node-in-tree-flag} follows from induction hypothesis \ref{is-node-in-tree-flag}.

\item
Suppose $v$ is active in $C$ and $s$ is any other step.  Then the truth of claim \ref{is-node-in-tree-flag} follows from induction hypothesis \ref{is-node-in-tree-flag}.
\end{itemize}
\item
Suppose that $v.update = \langle\mk,\&I\rangle$ in $C'$ and no child \CAS\ belonging to $I$ has occurred before $C'$.  Then, $v$ is active in $C$ since, immediately
after the child \CAS\ that makes $v$ active, $v$ is flagged for the dummy object.
We consider three cases.
\begin{itemize}
\item
Suppose $s$ is a successful mark \CAS\ on $v$.  Then this mark \CAS\ belongs to $I$ since $v.update = \langle\mk,\&I\rangle$ in $C'$.
Let $up$ be the value stored in $v.update$ in configuration $C$.
If $up.type =\flag$, then claim \ref{is-node-in-tree-mark1} follows from induction hypothesis \ref{is-node-in-tree-flag}.
Now suppose $up = \langle\mk,\&I'\rangle$ for some Info object $I'$.
Prior to creating $I$ at line \lref{create-info}, \func{Frozen}($up$) at line \lref{execute-check-frozen}
returned \FALSE.  
Thus, $I'.state$ was \abort.  By Lemma \ref{abort-commit-exclusive}, there is no child \CAS\ belonging to $I'$.
So, $v$ is in $T_i(C')$ by inductive hypothesis \ref{is-node-in-tree-mark1}.
\item
Suppose $s$ is a successful child \CAS.  This child \CAS\ must belong to some Info object $I' \neq I$, since we assumed that no child \CAS\ of $I$ occurs before $C'$.  
The argument that $v$ is in $T_i(C')$ is identical to the argument for the third case of \ref{is-node-in-tree-flag}, above.
\item 
Suppose  $s$ is any other step.  Then claim \ref{is-node-in-tree-mark1} follows from induction hypothesis \ref{is-node-in-tree-mark1}.
\end{itemize}
\item
Suppose that $v.update = \langle\mk,\&I\rangle$ in $C'$ and $i<I.seq$.
We consider four cases.
\begin{itemize}
\item
Suppose $s$ is a successful mark \CAS\ on $v$. 
The argument that $v$ is in $T_i(C')$ is identical to the argument for the first case of \ref{is-node-in-tree-mark1}, above.
\item
Suppose $s$ is a successful child \CAS\ that belongs to $I$.
By Corollary \ref{only-first-child-CAS}, there is no child \CAS\ belonging to $I$ before $C$.
By induction hypothesis \ref{is-node-in-tree-mark1}, $v$ is in $T_i(C)$.
By claim \ref{child-CAS-effect-low}, proved above, $T_i(C) = T_i(C')$.
So, $v$ is in $T_i(C')$. 
\item
Suppose $s$ is a successful child \CAS\ that belongs to some Info object $I'\neq I$. 
The argument that $v$ is in $T_i(C')$ is identical to the argument for the third case of \ref{is-node-in-tree-flag}, above.
\item 
Suppose  $s$ is any other step.  Then the truth of claim \ref{is-node-in-tree-mark2} follows from induction hypothesis \ref{is-node-in-tree-mark2}.
\end{itemize}
\end{enumerate}
\end{enumerate}
\end{proof}

\begin{corollary}
\label{no-new-ancestors}
Let $v$ be a node that is active in some configuration $C$. Then, for every $i \geq 0$, 
if $v$ is in the left (or right) subtree of a node $v'$ with key $k$ within tree $T_i(C')$ for some later configuration $C'$,
then $v$ was in the left (or right, respectively) subtree of a node with key $k$ within tree $T_i(C)$.
\end{corollary}
\begin{proof}
This follows immediately from Lemma \ref{child-CAS-super-lemma}.\ref{child-CAS-effect}.
\end{proof}

Given a binary tree (which may or may not be a BST), we define the \emph{search path
for a key $k$} to be the path that begins at the root and, at each node, passes
to the left or right child, depending on whether $k$ is less than the key in the node or not.

\begin{lemma}
\label{still-on-search-path}
If, for each $i \geq 0$, a node $v$ is on the search path for key $k$ in $T_i(C)$  for some configuration $C$
and is still in $T_i(C')$ for some later configuration $C'$, then $v$ is on the search path for $k$ in $T_i(C')$.
\end{lemma}
\begin{proof}
This follows immediately from Corollary \ref{no-new-ancestors}.
\end{proof}

\later{If the following lemma is only used once, then remove it and just explain why it is true where it is used.}

\begin{lemma}
\label{readchild-result-valid}
A call to \func{ReadChild}($p, \lft, seq$) returns the version-$seq$ left (or right) child of the node pointed to by $p$ at the time line~\lref{read-child} is executed
if \lft\ is \TRUE\ (or \FALSE, respectively).
\end{lemma}

\begin{proof}
This follows immediately from the fact that $prev$ fields of nodes never change (by Observation \ref{unchanging}).
\end{proof}

Whenever \func{Search}($k,seq$) reads a left (or right) child field  of a node $v$ on line~\lref{read-child} then we say that the \func{Search} {\em visits}  the 
version-$seq$ left (or right, respectively) child of $v$. (Notice that the time a node is visited is earlier than the time that local variable $\ell$ of \func{Search} points to this node.)  We also say that a \func{Search} {\em visits} the root when it executes line \lref{search-initialize}.

\begin{lemma}
\label{search-linearizable}
Consider any instance $S$ of \func{Search}($k, seq$) that terminates, and let $v_1, \ldots, v_k$ be the nodes visited by $S$ (in the order they are visited).
There exist configurations $C_1, C_2, \ldots, C_k$ such that 
\begin{enumerate}
\item\label{searchlin-start}
$C_1$ is after the search is invoked,
\item\label{searchlin-order}
for $i>1$, $C_{i-1}$ is before or equal to $C_i$,
\item \label{searchlin-onpath}
$v_i$ is on the search path for $k$ in $T_{seq}(C_i)$,
\item\label{searchlin-beforevisit}
$C_i$ is before the step where $S$ visits $v_i$, and
\item\label{searchlin-last}
$C_i$ is the last configuration that satisfies both (\ref{searchlin-onpath}) and (\ref{searchlin-beforevisit}).
\end{enumerate}
\end{lemma}
\begin{proof}
Since $v_1$ is the root node, which is visited when $S$ executes line \lref{search-initialize}, let $C_1$ be the configuration before $S$ executes line \lref{search-initialize}.  This satisfies all claims (including \ref{searchlin-order}, vacuously).

Let $1 < i \leq k$ and suppose $C_{i-1}$ has already been defined to satisfy all of the claims.
Let $C'$ be the configuration before $S$ visits $v_i$ by reading a child pointer of $v_{i-1}$.
Note that $C_{i-1}$ is before $C'$ by induction hypothesis \ref{searchlin-beforevisit}.
We first show that $v_i$ is on the search path for $k$ in $T_{seq}$ at some configuration between $C_{i-1}$ and $C'$ by considering two cases.

{\bf Case 1} ($v_{i-1}$ is in $T_{seq}(C')$).  Then, by  induction hypothesis \ref{searchlin-onpath} and Lemma \ref{still-on-search-path}, $v_{i-1}$ is on the search path for $k$ in $T_{seq}(C')$.
So, $v_i$ is also on the search path for $k$ in $T_{seq}(C')$.

{\bf Case 2} ($v_{i-1}$ is not in $T_{seq}(C')$).
Let $C''$ be the last configuration between $C_{i-1}$ and $C'$ when $v_{i-1}$ was in $T_{seq}(C'')$.
By Lemma \ref{still-on-search-path}, $v_{i-1}$ is on the search path for $k$ in $T_{seq}(C'')$.
The step after $C''$ must be a child \CAS\ that removes $v_{i-1}$ from $T_{seq}$.  By Lemma \ref{marking-permanent},
$v_{i-1}$ is marked at all times after $C''$.  By Lemma  \ref{only-first-child-CAS} and \ref{frozen-at-child-CAS}, the child pointers of $v_{i-1}$ are never changed after $C''$.
Since $prev$ pointers of nodes never change either, the version-$seq$ children of $v_{i-1}$ never change
after $C''$.  Thus, $v_{i}$ is already the version-$seq$ child of $v_{i-1}$ at configuration $C''$
since $v_i$ is the version-$seq$ child of $v_{i-1}$ at $C'$ after $C''$ by Lemma \ref{readchild-result-valid}.
Thus, $v_i$ is on the search path for $k$ in $T_{seq}(C'')$.

Thus, in either case, there is a configuration between $C_{i-1}$ and $S$'s visit to $v_i$ when $v_i$ is on the search path for $k$ in $T_{seq}$.  Let $C_i$ be the last such configuration.  The claims follow.
\end{proof}

\begin{invariant}
\label{search-path-invariant}
Let $C$ be any configuration and let $j\leq i$.  Suppose the search path for a key $k$ in $T_j(C)$ includes
a node $v$ and $v\in T_i(C)$.  Then the search path for $k$ in $T_i(C)$ also includes $v$.
\end{invariant}
\begin{proof}
The claim is true for the initial configuration $C_0$, since $T_j(C_0) = T_i(C_0)$.
We show that every step preserves the invariant.  The only step that changes a tree or search path is a successful child \CAS.
Consider a successful child \CAS\ belonging to some Info object $I$.
It changes a child pointer from $I.oldChild$ to $I.newChild$.  We consider three cases.

\begin{itemize}
\item
Suppose $I.newChild\rightarrow seq > i$.  Then by Lemma \ref{simple-inv}.\ref{info-inv}, neither $T_i$ nor $T_j$ change, so the invariant is preserved.

\item
Suppose $j\leq I.newChild\rightarrow seq \leq i$.
Let $C$ and $C'$ be the configurations before and after the successful child \CAS.

If the child \CAS\ is a dchild \CAS, then by Lemma \ref{child-CAS-super-lemma}, $T_j$ is not affected, while in $T_i$,
a parent $x$ and its children $y$ and leaf $z$ are replaced by a copy $y'$ of $y$, so that $x, y$ and $z$ are no longer in $T_i(C')$.
Thus, any search path that passed through $x$ in $T_i(C)$ will now instead pass through the new node $y'$ in $T_i(C')$.
If the search path continued to a child of $y$ in $T_i(C)$, it will continue to the same child of $y'$ in $T_i(C')$.
Thus, the invariant is preserved.

If the child \CAS\ is an ichild \CAS, then by Lemma \ref{child-CAS-super-lemma}, $T_j$ is not affected,
while in $T_i$, a leaf $x$ is replaced by an internal node with two leaf children.  The old leaf is no longer
in the tree $T_i(C')$.  Thus, all search paths in $T_i$ are unaffected, except those that pass through
 $x$, but since $x$ is not in $T_i(C)$, the invariant is still true for $C'$.

\item
Suppose $I.newChild\rightarrow seq \leq j$.  Then applies an identical change to both $T_i$ and $T_j$, so the invariant is preserved.
\end{itemize}
\end{proof}

\begin{invariant}
\label{bst-inv}
For every configuration $C$ and every integer $i \geq 0$, $T_i(C)$ is a BST.
\end{invariant}
\begin{proof}
The claim is true in the initial configuration.  The only steps that can modify $T_i$ are successful child \CAS\ steps, so we show that each successful child \CAS\ preserves the invariant.
Let $I$ be the Info object that this child \CAS\ belongs to and let $j=I.seq$.
If $i<j$ then the child \CAS\ does not affect $T_i$, by Lemma \ref{child-CAS-super-lemma}.\ref{child-CAS-effect-low}.
So suppose $i\geq j$.

First, consider a dchild \CAS.  By Lemma \ref{child-CAS-super-lemma}.\ref{dchild-CAS-effect-high}, the change to $T_i$ preserves the invariant.

Now, consider an ichild \CAS.   $I$ was created by an \func{Insert}($k$) operation. 
The change that this ichild \CAS\ can make to $T_i$ is described by Lemma \ref{child-CAS-super-lemma}.\ref{ichild-CAS-effect-high}: it replaces a leaf $l$ with key $k'$ by an internal node with two children whose keys are $k$ and $k'$.
By Lemma \ref{search-linearizable}, $l$ was on the search path for $k$ in $T_j$ in some configuration during the \func{Search}($k,j$) at line \lref{insert-search} of the \func{Insert}.
By Lemma \ref{frozen-at-child-CAS}, $l$ is marked for $I$ when the child \CAS\ occurs.
By Lemma \ref{child-CAS-super-lemma}.\ref{is-node-in-tree-mark1}, $l$ is still in $T_j$ in the configuration prior to the child \CAS.
By Lemma \ref{still-on-search-path}, $l$ is still on the search path for $k$ in $T_j$ in that configuration.
By Invariant \ref{search-path-invariant}, $l$ is also on the search path for $k$ in $T_i$ in that configuration.
Thus, the change to $T_j$, as described by Lemma \ref{child-CAS-super-lemma}.\ref{ichild-CAS-effect-high} preserves the BST invariant because the key $k$ is being inserted at the correct location in $T_j$.
\end{proof}

\subsubsection{Linearizability}
\label{sec:linearizability-argument}

Finally, we are ready to prove that the implementation is linearizable.
We do this by defining linearization points for all operations and proving Lemma \ref{DFI-correct},
which describes how the current state  of the data structure reflects the abstract set that would
be obtained by performing all of the operations that have been linearized so far
atomically at their linearization points.
This connection between the states of the actual data structure and the abstract set
also allows us to show that the results of all operations are consistent with this linearization.


We first show that \func{Help} returns an appropriate response that indicates whether the update being helped
has succeeded.

\begin{lemma}
\label{help-works}
Consider any call $H$ to \func{Help} that is called with a pointer to an Info object $I$.
\begin{enumerate}
\item
If $H$ returns \TRUE\ then there is a unique successful child \CAS\ that belongs to $I$, and that child \CAS\ occurs before $H$ terminates.
\item
If $H$ returns \FALSE\ then there is no successful child \CAS\ that belongs to $I$.
\item
If $H$ does not terminate then there is at most one successful child \CAS\ that belongs to $I$.
\end{enumerate}
\end{lemma}

\begin{proof}
\begin{enumerate}
\item
Suppose $H$ returns \TRUE.  Then, $I.state = \commit$ at line \lref{help-return}.  So
some call to \func{Help}($\&I$) performed a commit write at line \lref{commitWrite} prior to $H$'s execution
of line \lref{help-return}.
Prior to that, the same call to \func{Help} performed a child \CAS\ belonging to $I$.
By Lemma \ref{child-CAS-succeeds}, the first such child \CAS\ succeeds.
By Lemma \ref{only-first-child-CAS}, there is exactly on successful child \CAS\ belonging to $I$.

\item
Suppose $H$ returns \FALSE.  
By Lemma \ref{done-trying}, when $H$ reaches line \lref{help-return}, the $I.state$ must be \abort\ or \commit.
Since $H$ returns \FALSE, $I.state$ is \abort\ at line \lref{help-return}.
By Lemma \ref{abort-commit-exclusive}, there is no child CAS that belongs to $I$.

\item
This claim follows immediately from Lemma \ref{only-first-child-CAS}.
\end{enumerate}
\end{proof}

Next, we use the preceding Lemma to argue that each update returns an appropriate response, indicating
whether the update has had an effect on the data structure.

\begin{lemma}
\label{child-CAS-existence}
Consider any call $U$ to \func{Insert} or \func{Delete}.
\begin{enumerate}
\item
If $U$ does not terminate then there is at most one successful child \CAS\ that belongs to any Info object created by $U$.  If there is such a child \CAS, it belongs to the Info object created in the last iteration of $U$'s while loop.
\item
If $U$ returns \TRUE\ then there is exactly one successful child \CAS\ that belongs to any Info object created by $U$, and it belongs to the Info object created in the last iteration of $U$'s while loop.
\item
If $U$ returns \FALSE\ then there is no successful child \CAS\ that belongs to any Info object created by $U$.
\end{enumerate}
\end{lemma}

\begin{proof}
For each iteration of $U$'s while loop except the last, either \func{Execute} is not called or \func{Execute} returns \FALSE.
If \func{Execute} returns \FALSE, then either \func{Execute} does not perform the first
freezing \CAS\ successfully at line \lref{flagCAS1} or the call to \func{Help} returns \FALSE.
If the first freezing \CAS\ does not succeed, no process can call \func{Help} on the Info object created in this iteration of $U$.
If \func{Help} returns \FALSE, there is no child \CAS\ belonging to the Info object created in this iteration of $U$'s loop, by Lemma \ref{help-works}.
Thus, in all cases, there is no child \CAS\ belonging to an Info object created in this iteration of $U$'s loop.

The final iteration of $U$'s loop can create at most one Info object, which has at most one successful child \CAS, by Lemma \ref{only-first-child-CAS}.
This establishes claim (1) of the lemma.

If $U$ returns \TRUE, then $U$'s call to \func{Execute} on line \lref{insert-execute} or \lref{delete-execute}
returns true.  This means that the call to \func{Help} on line \lref{execute-help-self} of \func{Execute}
returns true.  By Lemma \ref{help-works}, there is exactly one successful child \CAS\ that belongs to the Info object
created in the final iteration of $U$'s loop.  This completes the proof of claim (2).

If $U$ returns \FALSE, then either \func{Execute} is not called at line \lref{insert-execute} or \lref{delete-execute}, or that call to \func{Execute} returns \FALSE.
By the same argument as in the first paragraph of this proof, there is no child \CAS\ associated with the Info
object created in the final iteration of $U$'s while loop.
\end{proof}

Next, we describe how operations of an execution are linearized.
For the remainder of the proof, we fix an execution $\alpha$.

If there is a successful child \CAS\ that belongs to an Info object $I$ created by an \func{Insert} or \func{Delete} operation,
we linearize the operation at the first freeze \CAS\ belonging to $I$ (at line \lref{flagCAS1}).
There is at most one such successful child \CAS, by Lemma \ref{child-CAS-existence}
and if such a child \CAS\ exists, it is preceded by a freezing \CAS, by Lemma \ref{frozen-at-child-CAS},
so this defines a unique linearization point for each update operation that has a successful child \CAS.
In particular, this defines a linearization point for every update operation that returns \TRUE\ and some that do not terminate, but it does not define a linearization point for any update that returns \FALSE,
by Lemma \ref{child-CAS-existence}. 

We linearize each \func{Insert} that returns \FALSE, each \func{Delete} that returns \FALSE\ and each \func{Find} that terminates in the operation's last call to \func{ValidateLeaf} at line \lref{insert-validate-leaf}, \lref{delete-validate-leaf} or \lref{find-validate-leaf}, respectively.  More specifically, we linearize 
the operation when $pupdate$ is read at line \lref{validate-reread} of that call to \func{ValidateLeaf}.
 
For each completed \Scan\ operation, we define its sequence number to be the value it reads from $Counter$ at line \lref{scan-getseq}.
We linearize 
every \Scan\ operation with sequence number $i$ at the step that the $Counter$ value changes from $i$ to $i+1$
with ties broken in an arbitrary way. 
Note that this step is well-defined and occurs during the execution interval of the \Scan:  after the \Scan\ reads $i$ from 
$Counter$, some process must increment $Counter$ from $i$ to $i+1$ no later than the \Scan's own increment at line \lref{scan-inc}.

In the following, we define an update operation to be imminent if its linearization point has occurred,
but it has not yet made the necessary change to the data structure.

\begin{definition}
\label{imminent}
An update operation is called \emph{imminent} in a configuration $C$ of execution $\alpha$ if, for some Info object $I$ created by the update,
\begin{itemize}
\item
there is a freezing \CAS\ belonging to $I$ before $C$,
\item
there is no child \CAS\ belonging to $I$ before $C$, and
\item
there is a child \CAS\ belonging to $I$ after $C$.
\end{itemize}
\end{definition}

The following lemma is a consequence of the way that update operations must freeze nodes in order to apply changes.

\begin{lemma}
\label{no-two-imminent}
In any configuration $C$, there cannot be two imminent updates with the same key.
\end{lemma}
\begin{proof}
To derive a contradiction, suppose there are two update operations $op_1$ and $op_2$ with the same key 
that are both imminent in $C$.
Let $I_1$ and $I_2$ be the two Info objects that satisfy definition \ref{imminent}.
Let $gp_1, p_1, l_1$ and $gp_2, p_2, l_2$ be the results of the last \func{Search} performed by the two operations
prior to creating $I_1$ and $I_2$, respectively.

$I_1.nodes$ includes $p_1$ and $I_2.nodes$ includes $p_2$.
By Lemma \ref{no-child-changes}, $l_1$ is the child of $p_1$ at configuration $C$.
Similarly, $l_2$ is the (same) child of $p_2$ at configuration $C$.
By Lemma \ref{search-linearizable}, $p_1$ and $l_1$ were all on the search path for $k$ in $T_\infty$ at some time before $C$.
By Lemma \ref{child-CAS-super-lemma}.\ref{is-node-in-tree}, they are still on the search path for $k$ in the configuration prior to the successful child \CAS\ of $I_1$.
So, by Lemma \ref{still-on-search-path}, they are on the search path for $k$ in $C$.
A similar argument shows that $p_2$ and $l_2$ are on the search path for $k$ in $C$.
So, $l_1=l_2$ and $p_1=p_2$.

Since $p_1$ appears in both $I_1$ and $I_2$ there must be a successful freezing \CAS\ belonging to each of $I_1$ and $I_2$ on this node, by Lemma \ref{frozen-at-child-CAS}.
Let $fcas_1$ and $fcas_2$ be the steps that freeze $p_1$ for $I_1$ and $I_2$, respectively.
Without loss of generality, assume $fcas_1$ occurs before $fcas_2$.
Then, $op_2$ reads a value $up$ from $p_1.update$ and stores the result in $I.oldUpdate$ {\it after} $fcas_1$;
otherwise $fcas_2$ would fail, by Lemma \ref{update-no-ABA}.
After $op_2$ reads this field, it gets the result \FALSE\ from \func{Frozen}$(up)$ at line \lref{execute-check-frozen} (otherwise the attempt would be aborted before $I_2$ is created at line \lref{create-info}).
Thus, $I_1.state$ must be \abort\ or \commit\ when \func{Frozen} checks this field.
By Lemma \ref{abort-commit-exclusive}, $I_1.state$ cannot be \abort\ because there is a child \CAS\ that belongs to $I_1$.
Thus, there is a commit write belonging to $I_1$ prior to $op_2$'s creation of $I_2$.
By the code, there is a child \CAS\ belonging to $I_1$ prior to the creation of $I_2$.
This contradicts the fact that the first child \CAS\ of $I_1$ occurs after $C$ but the first freezing \CAS\ belonging to $I_2$ occurs before $C$.
\end{proof}

The following lemma shows will be used to argue about the linearization point of a \func{Find} or an unsuccessful update operation, using the fact that \func{ValidateLeaf} has returned true.

\begin{lemma}
\label{validation-correct}
If a call \func{ValidateLeaf}($gp, p, l, k$) returns $\langle \TRUE, gpupdate, pupdate\rangle$ then in the configuration $C$ immediately before it reads $p.update$ at line \lref{validate-reread}, the following statements hold.
\begin{enumerate}
\item
Either ($k<p.key$ and $p.left = l$) or ($k\geq p.key$ and $p.right = l$).
\item
$p.update = pupdate$ and $pupdate$ is not frozen.
\item
If $p\neq Root$, either ($k<gp.key$ and $gp.left = p$) or ($k\geq gp.key$ and $gp.right = p$).
\item
If $p\neq Root$, $gp.update = gpupdate$ and $gpupdate$ is not frozen.
\end{enumerate}
\end{lemma}
\begin{proof}
Since \ValidateLeaf\ returns \TRUE, its calls to \ValidateLink\ return \TRUE.
\begin{enumerate}
\item
Consider the call to \ValidateLink\ at line \lref{validate-leaf-p} of \ValidateLeaf.
At line \lref{val-read-pupdate}, $p.update=pupdate$.  
Later, $p.update = pupdate$ at $C$.  By Lemma \ref{update-no-ABA}, $p.update$ was equal to $pupdate$ throughout that
period.  Node $p$ was not frozen at line \lref{val-check-frozen}, so no changes to $p$'s children occurred
between that time and $C$, by Lemma \ref{frozen-at-child-CAS}.  Claim (1) was true when line \lref{val-read-child} was performed during that interval,
so it is still true at $C$.
\item
Since \ValidateLeaf\ returns \TRUE, $p.update = pupdate$ when it is read at line \lref{validate-reread} just after configuration $C$.  Moreover, $pupdate$ was not frozen during the call to \ValidateLink\ at line \lref{validate-leaf-p} before $C$, so it is still not frozen in $C$, by Corollary \ref{irreversible}.
\item
Suppose $p\neq Root$.
Consider the call to \ValidateLink\ at line \lref{validate-gp-p} of \ValidateLeaf.
At line \lref{val-read-pupdate}, $gp.update=gpupdate$.  
Later, $gp.update = gpupdate$ at the read of $gp.update$ on line \lref{validate-reread}, which is after $C$.  By Lemma \ref{update-no-ABA}, $gp.update$ was equal to $gpupdate$ throughout that
period (including at $C$).  
Node $gp$ was not frozen at line \lref{val-check-frozen}, so no changes to $gp$'s children occurred
between that time and $C$, by Lemma \ref{frozen-at-child-CAS}.  Claim (3) was true when line \lref{val-read-child} was performed during that interval,
so it is still true at $C$.
\item
Before $C$, $gp.update = gpupdate$ at line \lref{val-read-pupdate} of the call to \ValidateLink\ on line \lref{validate-gp-p}.
Since \ValidateLeaf\ returns \TRUE, $gp.update = gpupdate$ when it is read at line \lref{validate-reread}  after configuration $C$.  
By Lemma \ref{update-no-ABA}, $gp.update = gpupdate$ at configuration $C$.
Moreover, $gpupdate$ was not frozen during the call to \ValidateLink\ at line \lref{validate-gp-p} before $C$, so it is still not frozen in $C$, by Corollary~\ref{irreversible}.
\end{enumerate}
\end{proof}

Now, we are ready to establish the connection between the state of the shared data structure and the abstract set that it represents (according to the operations that have been linearized so far).
For any configuration $C$ of execution $\alpha$, let 
\begin{eqnarray*}
L(C) & = & \{k : \mbox{there is a leaf of $T_\infty(C)$ with key } k\}\\
I_{ins}(C) & = & \{k : \mbox{there is an imminent \func{Insert}($k$) in } C\}\\
I_{del}(C) & = & \{ k : \mbox{there is an imminent \func{Delete}($k$) in } C\}\\
Q(C) &=& (L(C) \cup I_{ins}(C)) - I_{del}(C)
\end{eqnarray*}
Let $S(C)$ be the set of keys that would result if all update operations whose linearization points are before $C$ were performed atomically in the order of their linearization points.  

\begin{lemma}
\label{DFI-correct}
For all configurations $C$ in execution $\alpha$, 
\begin{enumerate}
\item
\label{key-inv}
$Q(C)=S(C) \cup \{\infty_1, \infty_2\}$, 
\item
\label{imminent-inserts}
$I_{ins}(C) \cap L(C) = \{\}$,
\item
\label{imminent-deletes}
$I_{del}(C) \subseteq L(C)$,
and
\item
\label{right-answers}
If a \func{Find}, \func{Insert} or \func{Delete} operation that terminates in $\alpha$
is linearized in the step after configuration $C$, the output it returns is the same
as if the operation were done atomically on a set in state $S(C)$.
\end{enumerate}
\end{lemma}
\begin{proof}
We prove the claim holds for all states and linearization points in a prefix of the execution, by induction on the length of the prefix.

For the base case, consider the prefix of 0 steps.
In the initial configuration $C$, we have $Q(C) = \{\infty_1,\infty_2\}$, $I_{ins}(C)=I_{del}(C) = S(C) = \{\}$.  There are no linearization points, so claim \ref{right-answers} holds vacuously.

Assume the claim is true for a prefix $\alpha'$.  We prove that it holds for $\alpha'\cdot s$ where $s$ is the next step of $\alpha$. Let $C$ and $C'$ be the configurations before and after $s$.
We consider several cases.

\begin{itemize}
\item
Suppose $s$ is the first freezing \CAS\ of an Info object  that has a child \CAS\ later in $\alpha$ and the Info object is created by an \func{Insert}($k$) operation.
This is the linearization point of the \func{Insert}.
So, $S(C') = S(C) \cup\{k\}$.
We have $L(C')=L(C)$, $I_{ins}(C') = I_{ins}(C)\cup \{k\}$ and $I_{del}(C')=I_{del}(C)$.
By Lemma \ref{no-two-imminent}, $k\notin I_{del}(C')$, so 
$Q(C') = Q(C) \cup \{k\}$.  Thus, $s$ preserves claims \ref{key-inv} and \ref{imminent-deletes}.

Let $gp, p$ and $l$ be the three nodes returned by the last \func{Search} at line \lref{insert-search} of the \func{Insert}.
By Lemma \ref{search-linearizable}, $p$ and its child $l$ were on the search path for $k$ in $T_\infty$ in some earlier configuration.  Since $p$ is flagged for the \func{Insert} in $C'$,
it follows from Lemma \ref{child-CAS-super-lemma}.\ref{is-node-in-tree-flag} that $p$ is still in the tree at $C'$.
By Lemma \ref{no-child-changes}, its child is still $l$ at $C'$.  Thus, $l$ is still on the search path for $k$ in $C'$, by Lemma \ref{still-on-search-path}.  Since the test on line \lref{insert-false} evaluated to \FALSE, $l.key \neq k$.  By Lemma \ref{bst-inv}, the tree $T_\infty(C')$ is a BST, so $k$ does not appear anywhere else in it.  Thus, $k\notin L(C') = L(C)$.  This ensures claim \ref{imminent-inserts} is preserved in $C'$.

By Lemma \ref{no-two-imminent} applied to $C'$, there is no imminent \func{Insert}($k$) in $C$.
So, $k\notin Q(C)$.  By the induction hypothesis, $k\notin S(C)$.  So, an \func{Insert}($k$) performed on a set in state $S(C)$ would return \TRUE.  By Lemma \ref{help-works} the \func{Insert}{$k$} linearized at $s$ also returns \TRUE, establishing claim~\ref{right-answers}.

\item
Suppose $s$ is the first freezing \CAS\ of an Info object  that has a child \CAS\ later in $\alpha$ and the Info object is created by a \func{Delete}($k$) operation.
This is the linearization point of the \func{Delete}.
So, $S(C') = S(C) - \{k\}$.
We have $L(C')=L(C)$, $I_{ins}(C') = I_{ins}(C)$ and $I_{del}(C')=I_{del}(C)\cup\{k\}$.
So $Q(C') = Q(C) - \{k\}$.  Thus, $s$ preserves claim \ref{key-inv} and \ref{imminent-inserts}.

Let $gp, p$ and $l$ be the three nodes returned by the last \func{Search} at line \lref{delete-search} of the \func{Delete}.
By Lemma \ref{search-linearizable}, these three nodes were on the search path for $k$ in $T_\infty$ in some earlier configuration.  The node $gp$ is flagged for the \func{Delete} in $C'$ and, by Lemma \ref{no-child-changes},
$p$ is still the child of $gp$ and $l$ is still the child of $p$ in $C'$.
It follows from Lemma \ref{child-CAS-super-lemma}.\ref{is-node-in-tree-flag} that $gp$ is still in the tree at $C'$.
Thus, $l$ is still on the search path for $k$ in $C'$, by Lemma \ref{still-on-search-path}.  Since the test on line \lref{delete-false} evaluated to \FALSE, $l.key = k$.   Thus, $k\in L(C') = L(C)$.  
This ensures claim \ref{imminent-deletes} is preserved in $C'$.

By Lemma \ref{no-two-imminent} applied to $C'$, there is no imminent \func{Delete}($k$) in $C$.
So, $k\in Q(C)$.  By the induction hypothesis, $k\in S(C)$.  So, a \func{Delete}($k$) performed on a set in state $S(C)$ would return \TRUE.  By Lemma \ref{help-works} the \func{Delete}{$k$} linearized at $s$ also returns \TRUE, establishing claim~\ref{right-answers}.

\item
Suppose $s$ is the first child \CAS\ of an \func{Insert}($k$) operation.
This is not the linearization point of any operation, so $S(C') = S(C)$.  Furthermore, claim \ref{right-answers} follows from the induction hypothesis.
By Lemma \ref{child-cas-lemma}, $L(C') = L(C) \cup \{k\}$.
By definition of imminent and Lemma \ref{no-two-imminent}, $I_{ins}(C') = I_{ins}(C) - \{k\}$.
Furthermore $I_{del}(C') = I_{del}(C)$.
So, $Q(C') = Q(C)$ and $S(C') = S(C)$, so claim \ref{key-inv}, \ref{imminent-inserts} and \ref{imminent-deletes} are preserved in $C'$. 

\item
Suppose $s$ is the first child \CAS\ of a \func{Delete}($k$) operation.
This is not the linearization point of any operation, so $S(C') = S(C)$.  Furthermore, claim \ref{right-answers} follows from the induction hypothesis.
By Lemma \ref{child-cas-lemma}, $L(C') = L(C) - \{k\}$.
By definition of imminent and Lemma \ref{no-two-imminent}, $I_{del}(C') = I_{del}(C) - \{k\}$.
Furthermore $I_{ins}(C') = I_{ins}(C)$.
So, $Q(C') = Q(C)$ and $S(C') = S(C)$, so claim \ref{key-inv}, \ref{imminent-inserts} and \ref{imminent-deletes} are preserved in $C'$. 

\item
Suppose $s$ is the linearization point of a \func{Find}($k$) that returns \TRUE\ or a \func{Insert}($k$) that returns \FALSE.
This linearization point is at the read of $p.update$ on line \lref{validate-reread} of the final \ValidateLeaf\ of the operation, which returns \TRUE.
By Lemma \ref{validation-correct}, $gp$ and $p$ are the grandparent and parent of $l$, and neither are frozen.
This means that $l$ is in $T_\infty(C)$ and hence $k\in L(C)$.
Moreover, there is no imminent \Delete($k$) (since then $gp$ would be frozen) so $k\notin I_{del}(C)$.
Hence, $k\in Q(C)$ and $k\in S(C)$ by the induction hypothesis.
So, $S(C') = S(C)$, since a \Find\ does not affect the abstract set and an \Insert($k$) would have no effect.
Also, $Q(C')=Q(C)$, so $S(C')=Q(C')$.
Moreover, a \func{Find}($k$) done atomically on the set $S(C)$ would return \TRUE\ and a \func{Insert}($k$) done atomically on the set $S(C)$ would return \FALSE.

\item
Suppose $s$ is the linearization point of a \func{Find}($k$) that returns \FALSE\ or a \func{Delete}($k$) that returns \FALSE.
This linearization point is at the read of $p.update$ on line \lref{validate-reread} of the final \ValidateLeaf\ of the operation, which returns \TRUE.
By Lemma \ref{validation-correct}, $gp$ and $p$ are the grandparent and parent of $l$, and neither are frozen.
This means that $l$ is in $T_\infty(C)$ and hence $k\notin L(C)$, since $T_\infty$ is a BST by Lemma \ref{bst-inv}
and $l$ is on the search path for $k$ in $T_\infty(C)$.
Moreover, 
there is no imminent \Insert($k$) (since then $p$ would be frozen) so $k\notin I_{ins}(C)$.
Hence, $k\notin Q(C)$ and $k\notin S(C)$ by the induction hypothesis.
So, $S(C') = S(C)$, since a \Find\ does not affect the abstract set and an \Delete($k$) would have no effect.
Also, $Q(C')=Q(C)$, so $S(C')=Q(C')$.
Moreover, a \func{Find}($k$) or \func{Delete}($k$) done atomically on the set $S(C)$ would return \FALSE.
\end{itemize} 
\end{proof}

Let $G$ be the directed graph consisting of all nodes, where there is an edge from node $u$ to node $v$ if $v$ was a child
of $u$ at some time during the execution.

\begin{lemma}
\label{acyclic}
$G$ is acyclic.
\end{lemma}
\begin{proof}
Lemma \ref{child-cas-lemma} implies that a child \CAS\ does not set up a new path between two nodes that were active before the child \CAS\ unless there was already a path between them.  So, each child \CAS\ preserves the truth of the lemma.
\end{proof}

The following Lemma states that any call to \func{ScanHelper} (that satisfies certain preconditions)
will output the right set of keys.  It will be used to prove that \Scan's output is correct.

\later{The following proof needs a lot of polishing and details filled in}
\begin{lemma}
\label{scan-correct}
Let $seq$ be an integer.
Suppose a completed invocation $S$ to \func{ScanHelper}($node, seq, a, b$) satisfies the following preconditions in the configuration $C$ before it is invoked.
\begin{itemize}
\item
$node$ is in $T_{seq}(C)$, 
\item
no proper ancestor of $node$ in $T_{seq}(C)$ is frozen in $C$ for a successful Info object with sequence number that is at most $seq$,
\item
$node$ is not permanently marked in $C$ for an Info object whose sequence number is at most $seq$, and
\item
$Counter > seq$ in $C$.
\end{itemize}
Let $C'$ be the configuration before $Counter$ is incremented from $seq$ to $seq+1$.
Then a key $k$ is in the set returned by $S$ iff
\begin{enumerate}
\item
\label{scan-inrange}
$k\in [a,b]$,
\item
\label{scan-onpath}
$node$ is on the search path for $k$ in $T_{seq}(C)$,
\item
\label{scan-inserted}
either $k$ appears in some leaf of the subtree of $T_{seq}(C')$ rooted at $node$ 
or there is a successful \func{Insert}($k$) with sequence number less than or equal to seq whose child CAS occurs after $C'$, and
\item
\label{scan-notdeleted}
there is no successful \func{Delete}($k$)
 with sequence number less than or equal to seq whose child CAS occurs after $C'$.
\end{enumerate}
\end{lemma}

\begin{proof}
Consider the subgraph $G_{seq}$ of $G$ consisting of nodes whose sequence numbers are less than $seq$.
$G_{seq}$ is finite since $Counter > seq$ at all times after $C_{seq}$, so only finitely many updates have sequence number at most $seq$.
By Lemma \ref{acyclic}, $G_{seq}$ is acyclic.
So, we  prove the claim by induction on the length of the longest path from $node$ to a sink in $G_{seq}$.

{\bf Base Case:}  If $node$ is a sink in $G_{seq}$, then it is a leaf node.

($\Rightarrow$):  Suppose $S$ returns $\{k\}$.  
By line \lref{scanhelper-leaf}, $k$ is the key of $node$ and $k\in [a,b]$.
So claim \ref{scan-inrange} is satisfied.
Since $node$ is in $T_{seq}(C)$ and $T_{seq}(C)$ is a BST by Lemma \ref{bst-inv}, $node$ is on the search path for $k$ in $T_{seq}(C)$, so claim \ref{scan-onpath} is satisfied.

Case 1:  If $T_{seq}(C')$ contains a leaf with key $k$, claim \ref{scan-inserted} is satisfied.  
If there is a \func{Delete}($k$) with sequence number at most $seq$ that is imminent in $C'$, then the child \CAS\ must be completed before $C$ since no proper ancestor of $node$ is frozen in $C$ for the \Delete; but $k$ cannot be re-inserted into $T_{seq}$ after $C'$, due to Lemma \ref{no-two-imminent} applied to configuration $C'$, contradicting the assumption that $node$ is in $T_{seq}(C)$ and contains $k$.  Thus, claim \ref{scan-notdeleted} is satisfied.

Case 2:  If $T_{seq}(C')$ does not contain a leaf with key $k$, (since $C$ is after $C'$) there must have been an \Insert($k$) that added a leaf with key $k$ to $T_{seq}$.  That insertion must have sequence number at most $seq$ (since otherwise it would not change $T_{seq}$, by Lemma \ref{child-CAS-super-lemma}).
Thus, claim \ref{scan-inserted} is satisfied.  Moreover, by Lemma \ref{no-two-imminent}, there cannot be
a \func{Delete}($k$) with sequence number at most $seq$ whose child \CAS\ occurs after $C'$.  Thus, claim \ref{scan-notdeleted} holds.

($\Leftarrow$):
Assume statments \ref{scan-inrange} to \ref{scan-notdeleted} are true for some key $k$.  We must show that $k$ is the key of $node$ (and hence is returned by $S$, since $k\in [a,b]$ by statement \ref{scan-inrange}).
We argue that $k$ is in a leaf of $T_{seq}(C)$.
By statement \ref{scan-inserted}, we can consider two cases.

Case 1:  If $k$ is in a leaf of the subtree of $T_{seq}(C')$ rooted at $node$, then $k$ is the key of $node$ since $node$ is a leaf.  By statement \ref{scan-notdeleted}, $k$ is a leaf of $T_{seq}(C)$.  \later{Could explain this in more detail using Lemma \ref{child-CAS-super-lemma}.}

Case 2:  If there is a successful \func{Insert}($k$) with sequence number at most $seq$ whose child \CAS\ occurs after $C'$.  Then, its child \CAS\ must occur before $C$ (because no ancestor of $node$ in $T_{seq}(C)$ is frozen for the \Insert).  So, by statement \ref{scan-notdeleted}, $k$ is in a leaf of $T_{seq}(C)$.

In either case, $T_{seq}(C)$ contains a leaf with key $k$.  Since $node$ is a leaf on the search path for $k$ of $T_{seq}(C)$, and $T_{seq}(C)$ is a BST by Lemma \ref{bst-inv}, $node$ must contain $k$.

{\bf Induction Step}:  Now suppose $node$ is an Internal node.  Assume the claim is true for calls to \func{ScanHelper} on nodes that are successors of $node$ in $G_{seq}$.  We prove that it is true for a call on $node$.

First, we argue that the recursive calls to \func{ScanHelper} satisfy the conditions of the lemma, so that we can apply the induction hypothesis to them.
Let $S_1$ be a recursive call to \func{ScanHelper} inside $S$ at line \lref{scanhelper-recursive1} to \lref{scanhelper-recursive4}.  Let $C_1$ be the configuration before $S_1$ is invoked.  Let $node_1$ be the node argument of $S_1$.
By hypothesis, none of $node$'s proper ancestors in $T_{seq}(C)$ are frozen with an Info object whose sequence number is less than or equal to $seq$ in $C$.  By handshaking, 
no update with sequence number at most $seq$ can freeze its first node after $C$ and succeed.
So by Lemma \ref{child-CAS-super-lemma}, the path in $T_{seq}$ from the root to $node$ never changes after $C$.
At some time during line \lref{scanhelper-help}, $node$ is not frozen for an in-progress Info object, by Lemma \ref{done-trying}.
So 
$node$'s version-$seq$ children do not change after this, and 
at configuration $C_1$ $node$ is in $T_{seq}(C_1)$ and $node_1$ is $node$'s version-$seq$ child,
so $node_1$ is also in $T_{seq}(C_1)$, as required.

Each proper ancestor of $node$ was not frozen in $C$ for a successful Info object with sequence number at most $seq$.
If any of those ancestors became frozen after $C$ with an Info object with sequence number at most $seq$, then
that Info object is doomed to abort due to handshaking.
Line \lref{scanhelper-help} ensures $node$ is not temporarily frozen (i.e., for an in-progress Info object) with sequence number at most $seq$, and handshaking ensures that it will never become so afterwards.
Since none of $node$'s ancestors is temporarily flagged in $C$ (with a sequence number at most $seq$) and $node$ is not 
permanently marked in $C$, it follows  that $node$ never gets permanently marked after $C$ by an Info object with sequence number at most $seq$.  \later{Probably need a lemma for previous statement}

Similarly, because none of $node_1$'s ancestors is flagged at $C_1$ by an Info object with sequence number at most $seq$, $node_1$ cannot be permanently marked by an Info object with sequence number at most $seq$ at $C_1$.

This completes the proof that the conditions of the Lemma are met for the recursive calls to \ScanHelper, so we can apply the induction hypothesis to them.

($\Rightarrow$):
Suppose $k$ is returned by $S$.  
We must prove that the 4 numbered claims are true for $k$.
The key $k$ is returned by one of the recursive calls $S'$ on line \lref{scanhelper-recursive1}--\lref{scanhelper-recursive4}.
Since $S'$ returns $k$, $k\in[a,b]$ by the induction hypothesis, so claim \ref{scan-inrange} is satisfied.
By the induction hypothesis, the version-$seq$ child of $node$ upon which $S'$ is called is on the search path for $k$
in $T_{seq}$ so $node is too$.
Similarly, claims \ref{scan-inserted} and \ref{scan-notdeleted} follow from the fact that they are satisfied for the recursive call $S'$.
 
($\Leftarrow$):  
Now suppose $k$ is some key that satisfies claims \ref{scan-inrange} to \ref{scan-notdeleted}.
If $k< node.key$, the four claims are satisfied for the  version-$seq$ left child of $node$,
and there is a recursive call on that child in line \lref{scanhelper-recursive2} or \lref{scanhelper-recursive4}, since $a\leq k < node.key$.
If $k\geq node.key$, the four claims are satisfied for the  version-$seq$ right child of $node$,
and there is a recursive call on that child in line \lref{scanhelper-recursive1} or \lref{scanhelper-recursive3},
since $b\geq k \geq node.key$.
Thus, one of the recursive calls returns $k$, and so does $S$.
\end{proof}

\begin{theorem}
The implementation is linearizable.
\end{theorem}

\begin{proof}
It follows from Lemma \ref{DFI-correct} and \ref{scan-correct} that each terminated operation returns the same value that it would if operations were performed atomically in the linearization ordering.
\end{proof}

\subsubsection{Progress}

The remaining results show that \Scan s are wait-free and all other operations are non-blocking.

\begin{lemma}
\label{readchild-wait-free}
Calls to \func{ReadChild} are wait-free.
\end{lemma}
\begin{proof}
Whenever a node is created, its $prev$ pointer is set to a node that already exists.
Thus, there can be no cycles among $prev$ pointers.
\end{proof}

\begin{theorem}
\label{scan-wf}
\Scan s are wait-free.
\end{theorem}
\begin{proof}
Let $\ell\geq 0$.
We prove that no call to \func{ScanHelper} with parameter $seq = \ell$ can take infinitely many steps.
Let $G_\ell$ be the subgraph of $G$ consisting of nodes whose $seq$ field is equal to $\ell$.
Note that $G_\ell$ is acyclic since $G$ is acyclic and finite, since the \Scan\ increments
$Counter$ from $\ell$ to $\ell+1$ and only nodes created by iterations of the while loops of update operations
that read $Counter$ before this increment can belong to $G_\ell$.

We prove the claim by induction on the maximum length of any path from $node$ to a sink of $G_\ell$:

Base case:  if $node$ is a sink of $G_\ell$, then it must be a leaf, so termination is immediate.

Inductive step:  \func{ScanHelper}($node, \ell, a,b$) calls \func{ScanHelper} on nodes that are successors of $node$ in $G_\ell$, which terminates by the induction hypothesis, and \func{ReadChild}, which terminates by Lemma \ref{readchild-wait-free}.

Then the claim follows, since \Scan\ just calls \func{ScanHelper}.
\end{proof}

\begin{theorem}
\label{lock-freedom}
The implementation is non-blocking.
\end{theorem}
\begin{proof}
To derive a contradiction, suppose there is an infinite execution where only a finite number of operations terminate.
Eventually, no more \Scan\ operations take steps, by Lemma \ref{scan-wf}, so the $Counter$ variable
stops changing.  Let $\ell$ be the final value of $Counter$.
Since there is at most one successful child \CAS\ belonging to each update operation,
there is a point in the execution after which there are no more changes to child pointers.

Suppose there is at least one update that takes infinitely many steps.
Let $O$ be the set of update operations that each take infinitely many steps without terminating.
Beyond some point, each \func{Search} performed by an operation in $O$ repeatedly returns the same three nodes $gp, p$ and $l$.
If $gp$ or $p$ is frozen, the operation calls \func{Help} on the Info object causing that Info object's state
to become \abort\ or \commit, by Lemma \ref{help-works}.
So, eventually these three nodes can be frozen for updates in $O$.
Consider a node $v$ in $G$ that is the $p$ node of some \func{Insert} in $O$ or the $gp$ node of some \func{Delete} in $O$ such that no other such node is reachable from $v$.  (Such a $v$ exists, since $G$ is acyclic and finite.)
One of the operations in $O$ will eventually sucessfully perform its first freeze \CAS\ on $v$, and then
no other operation can prevent it from freezing the rest of its nodes, so the operation will terminate, a contradiction.

Now suppose there is no update that takes infinitely many steps.  So, the operations that run forever are all
\func{Find} operations.  Let $O$ be the set of these operations.
Beyond some point, each \func{Search} performed by a \func{Find} in $O$ will repeatedly return the same $gp, p$ and $l$.  Due to helping, these nodes will eventually be unfrozen, so the \func{ValidateLeaf} called by \func{Find}
will return \TRUE\ and the \func{Find} will terminate, which is again a contradiction.
\end{proof}



\section{Open Questions}
\label{sec:discussion}

\vspace*{-2mm}

\later{mention garbage collection?}

\ignore{
We have presented a BST implementation that provides 
the first wait-free algorithm for \Scan\ 
on top of tree data structures, together
with non-blocking algorithms for \Insert, \Delete\
and \Find.
}

We believe that our approach can be generalized to work
on many other concurrent data structures. 
Could it be used, for example, to provide \Scan s for Natarajan and Mittal's 
implementation of a non-blocking leaf-oriented BST~\cite{NM14},
which records information about ongoing operations
in the tree edges they modify?  
Or with Natarajan {\em et al.}'s wait-free implementation of a red-black tree~\cite{NSM13}, which is based on the framework of \cite{TL94}?
More generally, could we design a general
 technique similar to~\cite{BER13,BER14}
to support wait-free partial \func{Scans} on top of 
any concurrent tree data structure?


\ignore{
\begin{itemize}
\item FIND and unsuccessfull Inserts and Deletes do not have to restart; they may continue from the point that they are;
we need a ggp pointer or a stack and that's it.

\item
if you don't like putting a flag/mark bit in same word as info object pointer, can do it separately like in LLX/SCX paper.  (for Java people)

\item
Optimization:  \func{ValidateLeaf} checks if gp or p are frozen and if so helps them.  So Execute doesn't really have to repeat this check for those nodes.  (In fact, Execute really only has to do this for the sibling node in a deletion.)  But we left it like this for generalization to other tree operations.
\end{itemize}
}

\vspace*{.2cm}

\noindent{\bf Acknowledgements}
A part of this work was done while Eric Ruppert was visiting FORTH ICS and the University of Crete.
Financial support was provided by the Natural Sciences and Engineering Research Council of Canada and by the European Commission under the Horizon 2020 Framework Programme for Research and Innovation through the EuroExa project (754337) and the HiPEAC Network of Excellence.

\newpage
{\small
\bibliographystyle{abbrv}

}



\end{document}